\def\eiso{E_{\rm iso}}
\def\egamma{E_{\gamma}}
\def\ep{E_{\rm peak}}
\def\epo{E^{\rm obs}_{\rm peak}}
\def\eps{E^{\rm src}_{\rm peak}}
\def\lpiso{L^{\rm peak}_{\rm iso}}

\documentclass[12pt,preprint]{aastex}
\usepackage{color}

\slugcomment{Not to appear in Nonlearned J., 45.}

\shorttitle{$\Gamma$-$\ep$ relation}
\shortauthors{Sakamoto et al.}

\begin{document}


\title{$\ep$ estimator for Gamma-Ray Bursts Observed by the 
{\it Swift} Burst Alert Telescope}


\author{T. Sakamoto\altaffilmark{1,2,3}, 
G. Sato\altaffilmark{8},
L. Barbier\altaffilmark{3}, 
S. D. Barthelmy\altaffilmark{3}, 
J. R. Cummings\altaffilmark{1,2,3}, 
E. E. Fenimore\altaffilmark{5},
N. Gehrels\altaffilmark{3}, 
D. Hullinger\altaffilmark{11}, 
H. A. Krimm\altaffilmark{1,7,3}, 
D. Q. Lamb\altaffilmark{9},
C. B. Markwardt\altaffilmark{1,6,3},
D. M. Palmer\altaffilmark{5},
A. M. Parsons\altaffilmark{3},
M. Stamatikos\altaffilmark{4,3},
J. Tueller\altaffilmark{3},
T. N. Ukwatta\altaffilmark{12,3}
}

\altaffiltext{1}{Center for Research and Exploration in Space Science 
and Technology (CRESST), NASA Goddard Space Flight Center, Greenbelt, MD 
20771}
\altaffiltext{2}{Joint Center for Astrophysics, University of Maryland, 
        Baltimore County, 1000 Hilltop Circle, Baltimore, MD 21250}
\altaffiltext{3}{NASA Goddard Space Flight Center, Greenbelt, MD 20771}
\altaffiltext{4}{Oak Ridge Associated Universities, P.O. Box 117, 
 Oak Ridge, Tennessee 37831.}
\altaffiltext{5}{Los Alamos National Laboratory, P.O. Box 1663, Los
Alamos, NM, 87545.}
\altaffiltext{6}{Department of Astronomy, University of Maryland, 
	College Park, MD 20742.}
\altaffiltext{7}{Universities Space Research Association, 10211 Wincopin 
	Circle, Suite 500, Columbia, MD 21044.} 
\altaffiltext{8}{Institute of Space and Astronautical Science, 
JAXA, Kanagawa 229-8510, Japan.}
\altaffiltext{9}{Department of Astronomy and Astrophysics, University of 
Chicago, Chicago, IL, 60637.}
\altaffiltext{10}{Joint Center for Astrophysics, University of Maryland, Baltimore County, 
1000 Hilltop Circle, Baltimore, MD 21250}
\altaffiltext{11}{Moxtek, Inc., 452 West 1260 North, Orem, UT 84057}
\altaffiltext{12}{Department of Physics, The George Washington University, 
Washington, D.C. 20052}


\begin{abstract}
We report a correlation based on a spectral simulation study 
of the prompt emission spectra of 
gamma-ray bursts (GRBs) detected by the {\it Swift} Burst Alert 
Telescope (BAT).  
The correlation is between the $\ep$ energy, which is 
the peak energy in the $\nu$F$_{\nu}$ spectrum, and the photon index 
($\Gamma$) derived from a 
simple power-law model.  The $\ep$ - $\Gamma$ relation, assuming 
the typical smoothly broken power-law spectrum of GRBs, is 
$\log \ep = 3.258 - 0.829\,\Gamma$ ($1.3 \leq \Gamma \leq 2.3$).
We take into account not only a range of $\ep$ 
energies and fluences, but also distributions for both the 
low-energy photon index and the high-energy photon index in the smoothly 
broken power-law model.  The distribution of burst durations in the BAT 
GRB sample is also included in the simulation.  
Our correlation is consistent with the 
index observed by BAT and $\ep$ measured by the BAT, and by other 
GRB instruments.  
Since about 85\% of GRBs observed by the BAT are acceptably fit 
with the simple power-law model because of the relatively narrow energy 
range of the BAT, this relationship can be used to estimate $\ep$ 
when it is located within the BAT energy range.  
\end{abstract}





\section{Introduction}

One of the fundamental characteristics of the prompt emission of 
gamma-ray bursts (GRB) is $\ep$, which is the peak energy 
in the $\nu$F$_{\nu}$ spectrum.  According to {\it Beppo}SAX and {\it HETE-2}  
observations, $\ep$ for GRBs is widely spread from a few keV 
to the MeV range as a single distribution \citep{kippen_xrf_astroph,sakamoto2005}.  
This broad single $\ep$ distribution strengthens the argument 
that these bursts arise from the same origin.  
Based on this observational evidence, there are
several works which try to understand a unified picture of GRBs.  For
instance, 
{\it the off-axis jet model} \citep{yamazaki2004,toma2005},
{\it the structured jet model} \citep{rossi2002,zhang2002,zhang2004}, 
and 
{\it the variable jet opening angle model} \citep{lamb2005} are the
popular unified jet models.  On the other hand, there are 
theoretical 
models to explain the broad $\ep$ distribution in the frame work of 
the internal shock model 
\citep{mes2002,mochkovitch2003,barraud2005} and the external shock model 
\citep{dermer1999,huang2002,dermer2003}.  

There are several important empirical relationships proposed based 
on the $\ep$ energy.  One of the most cited relationships is the 
correlation between $\ep$ in the GRB rest frame ($\eps$) and 
the isotropic radiated energy ($\eiso$), the so called the $\eps$-$\eiso$ (Amati) 
relation \citep{amati2002,amati2003}.  Since this relation is extended 
down to X-ray flashes \citep{sakamoto2004,sakamoto2006}, the dynamic range 
of this relation is $\sim$3 orders of magnitude in $\eps$ and $\sim$5 order 
of magnitude in $\eiso$.  The second correlation is between the $\eps$ 
energy and the collimation-corrected energy ($\egamma$), the so called 
$\eps$-$\egamma$ (Ghirlanda) relation \citep{ghirlanda2004}.  
According to \citet{ghirlanda2004}, this relation has much tighter 
correlation than the $\eps$-$\eiso$ relation.  \citet{liang2005} 
investigated a similar relationship, but without using $\egamma$ 
which is heavily dependent on the calculation of the jet opening angle.  
They found a good correlation between $\eps$, $\eiso$, and the 
achromatic break time in the afterglow light curve (t$_{jet}$).  
The third relationship is between $\eps$ and the isotropic peak luminosity 
($\lpiso$), the so called the $\eps$-$\lpiso$ (Yonetoku) relation 
\citep{yonetoku2004}.  The latest fourth relationship is between 
$\lpiso$, $\eps$, and the time scale of the brightest 45 per cent of 
the background subtracted counts in the light curve of the prompt 
emission \citep{firmani2006}.  
If these relationships are valid, they must be related to the fundamental 
physics of GRBs.  Thus, $\eps$ energy provides us fruitful knowledge 
about the characteristics of the prompt emission of GRBs.  Furthermore, 
knowing the $\epo$ energy is crucial to calculating the bolometric fluence 
which reflects the
total radiated energy in the prompt emission.  

After the launch of $Swift$ \citep{gehrels2004} in 2004, the Burst Alert 
Telescope (BAT; \citet{barthelmy2005}) has observed about 100 
GRBs per year.  In about half of the GRBs, 
the $\ep$ energies are very likely to be within the BAT energy range 
\citep{sakamoto2007}.  However, due to the relatively narrow 
energy band of the BAT (15-150 keV in the background subtracted spectrum using the 
mask modulation), the BAT has a difficulty in determining $\epo$.  Our
purpose of this study is to find a way to estimate $\epo$ when it lies 
within the BAT energy range.  

Here, we report a good correlation between the photon power-law index derived 
from a simple power-law model and $\ep$ based on the spectral simulation study.  
We use a sample of 31 long BAT GRBs that are well fitted with the power-law times 
exponential cutoff model, and also 26 GRBs observed by other GRB instruments 
concurrent with the BAT to confirm our correlation.  Our correlation 
provides an estimate for $\ep$ from the photon index in a simple power-law 
fit at the range from 1.3 to 2.3.  We also calculated the 1$\sigma$ confidence 
level of the estimated $\ep$ of our correlation.  

\section{BAT spectral simulation}
\begin{figure}
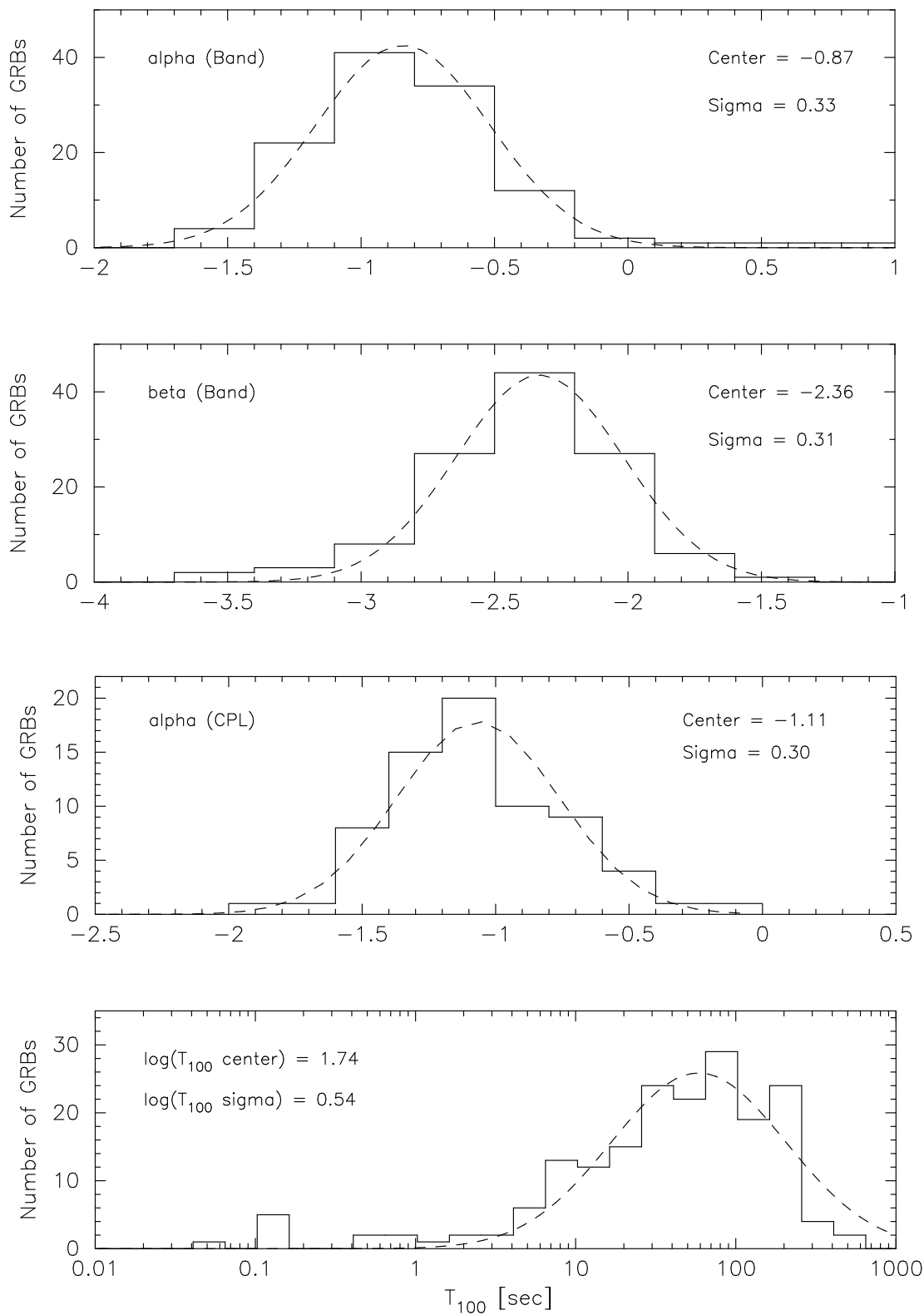

\centerline{
\includegraphics[scale=0.6,angle=-90]{f1a.eps}}
\vspace{1cm}
\centerline{
\hspace{0.3cm}
\includegraphics[scale=0.6,angle=-90]{f1b.eps}}
\caption{Input parameters in the spectral simulations.  The distribution of the low energy photon 
index $\alpha$, the high energy photon index $\beta$ in the Band function, the low energy photon 
index $\alpha$ in a CPL model from the BATSE GRB sample, 
and the BAT T$_{100}$ duration from top to bottom, respectively.  The dotted line represents the best fit in a 
gaussian.}
\label{sim_input_paramters}
\end{figure}

Because of the systematic difference in the spectral parameters based on 
the assumption of the spectral model \citep{band1993}, we decided to perform 
the simulations for two typical GRB spectral models as input spectra: 
the smoothly broken power-law model (Band function; \citet{band1993})
\footnote{dN/dE = K$_{1}$E$^{\Gamma_{1}} \exp[-\rm{E}(2+\Gamma_{1})/\ep]$ if
E $< (\Gamma_{1} - \Gamma_{2}) \ep/(2+\Gamma_{1})$ and
dN/dE = K$_{2}$E$^{\Gamma_{2}}$ if $E \geq
(\Gamma_{1} - \Gamma_{2}) \ep/(2+\Gamma_{1})$}
and a power-law times exponential cutoff 
model\footnote{dN/dE $\sim$ E$^{\alpha} \exp(-(2+\alpha)\,\rm{E}/\ep$)} (CPL) model.  
We fit the low-energy photon index, $\alpha$, and high-energy photon index, 
$\beta$, of 124 samples of the Band function fit (``BAND'' in their notation) 
in Table 9 of \citet{kaneko2006} by the normal distribution.  We obtained 
$\alpha$ of $-0.87$ with $\sigma$ of 0.33 and $\beta$ of $-2.36$ with $\sigma$ 
of 0.31.  
Note that we are not excluding the case of $\beta > -2$ in our simulations because 
two reports \citep[e.g.,][]{rsato2005,kaneko2006} show fits with $\beta > -2$ in 
both time-averaged and time-resolved burst spectra.  
However, the fraction of simulated spectra with $\beta > -2$ is 
only 13\% of the total.
Similarly, for a CPL model, we fit the low-energy photon index, 
$\alpha_{\rm CPL}$, for the sample in Table 9 of \citet{kaneko2006} (``COMP'' 
in their notation; 67 samples) by the normal distribution.  We found $\alpha_{\rm CPL}$ of $-1.11$ 
with $\sigma$ of 0.30 (see Figure \ref{sim_input_paramters}).  These $\alpha$, 
$\beta$, and $\alpha_{\rm CPL}$ distributions are used in our spectral simulation.  

In our simulations, $\ep$ varies from 1.4 keV to 1210 keV in a logarithmic 
scale.  The 15-150 keV fluence varies from 5 $\times$ 10$^{-8}$ to 5 $\times$
10$^{-5}$ ergs cm$^{-2}$ in a logarithmic scale.  The fluence range is determined 
based on the BAT observations (BAT1 catalog; \citet{sakamoto2007}).  
The simulation used 20 values for fluence and 70 values for $\ep$.  
The exposure time of the spectrum is the best fit log-normal 
distribution of the BAT T$_{100}$ duration\footnote{The duration includes 
from 0 to 100\% of the GRB fluence.} reported in the BAT1 catalog\footnote{The duration 
between tstart and tstop time of the fluence table.} 
(See the bottom panel of Figure \ref{sim_input_paramters}).  
The normalization of the input spectrum is calculated to be the 
input fluence value.  The spectral simulations are performed 1000 times 
for each grid point.  The background is included in the simulation using the 
spectrum created from the event data of the false BAT trigger 180931.  Since 
the background is subtracted using the mask modulation, the 
exposure time of the background spectrum is set as the same as the
duration of the foreground spectrum.  Four incident angles, 
on-axis (0$^{\circ}$), 15$^{\circ}$, 30$^{\circ}$, and 50$^{\circ}$ off-axis, 
are simulated independently.  The simulated spectra 
are fitted from 14 keV to 150 keV with a simple power-law
 model\footnote{dN/dE $\sim$ $E^{-\Gamma}$} (PL), a CPL, and 
the Band function.  {\tt Xspec} 11.3.2 was used in both creating and fitting 
the simulated spectra.  

\begin{figure}
\centerline{
\includegraphics[scale=0.5]{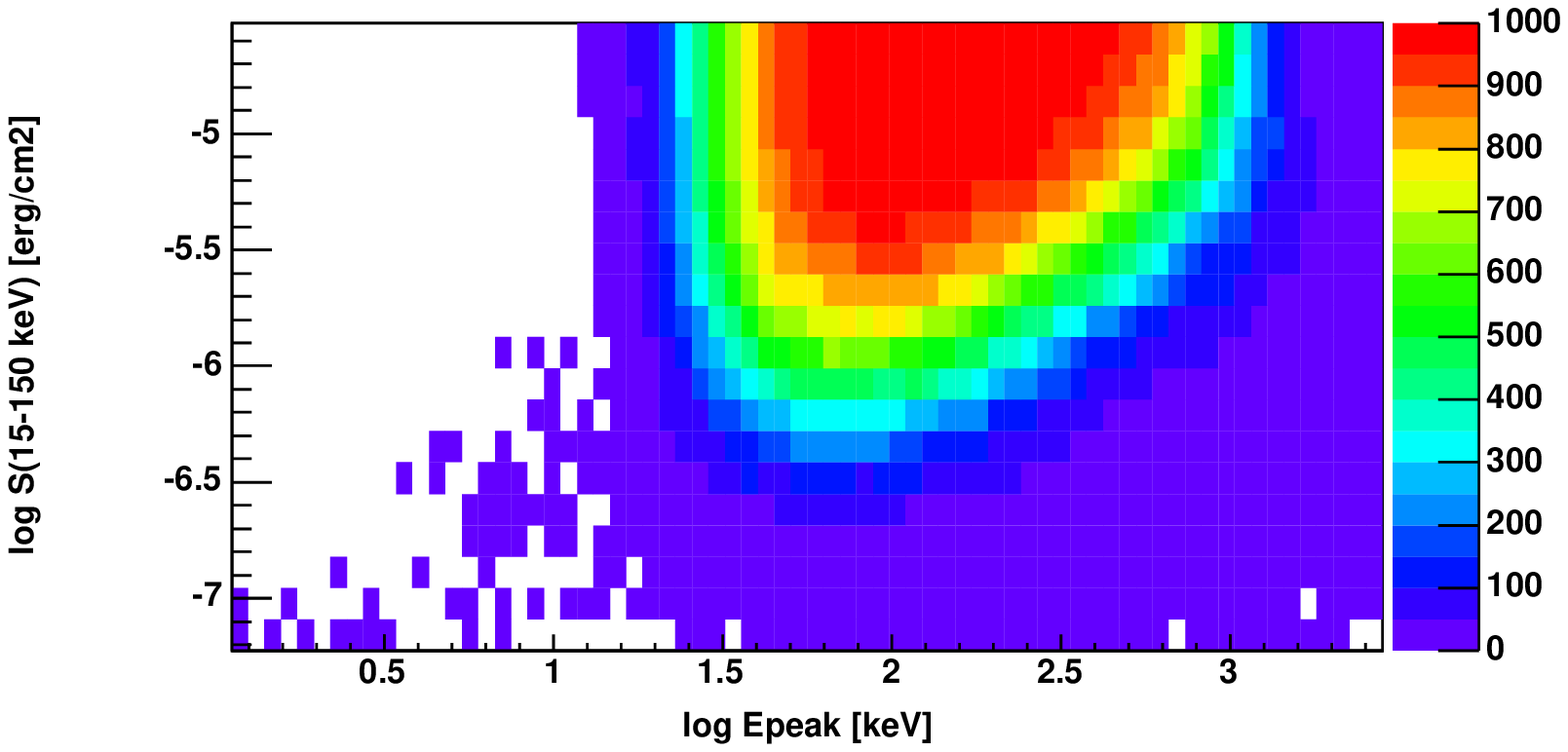}
\hspace{-1.0cm}
\includegraphics[scale=0.5]{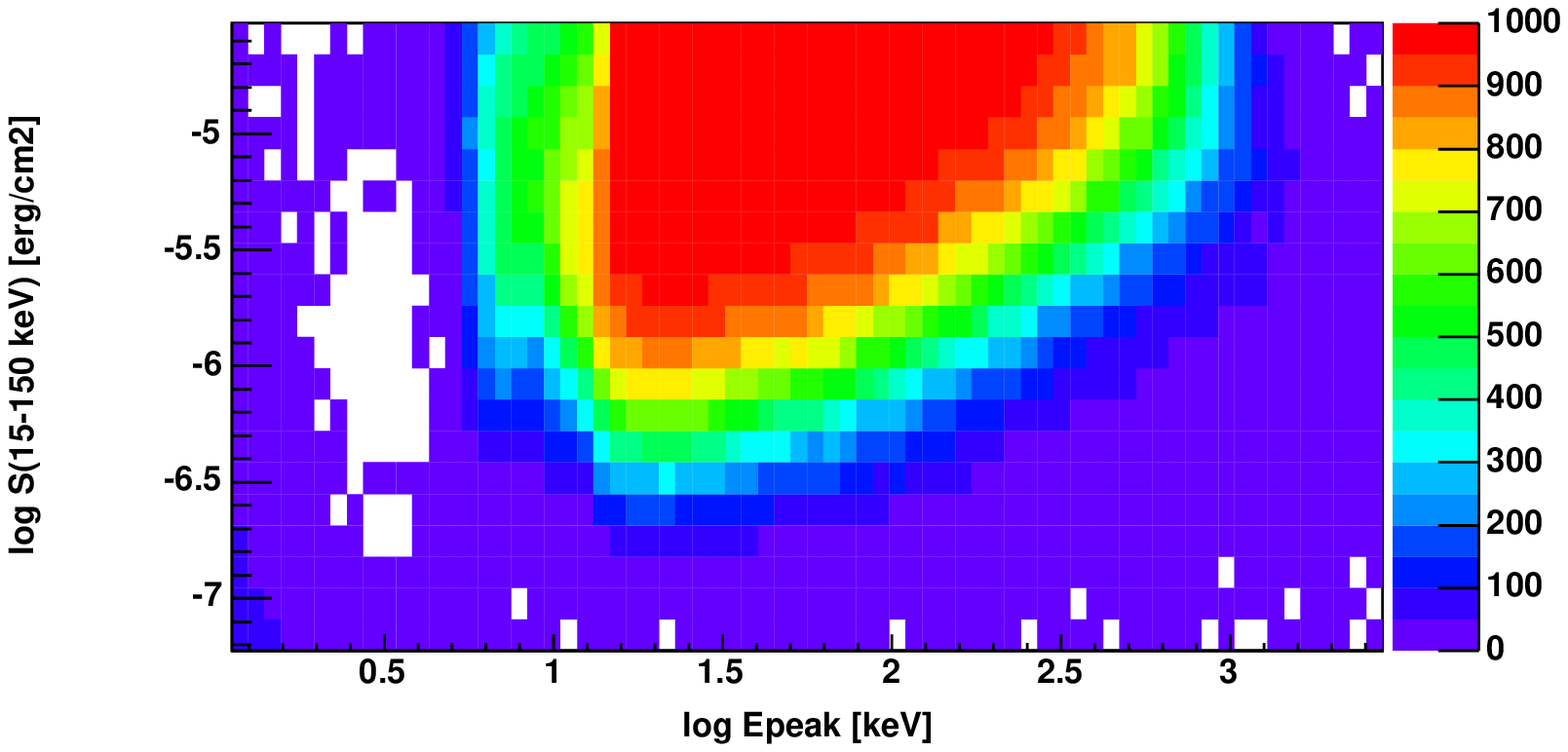}}
\centerline{
\includegraphics[scale=0.5]{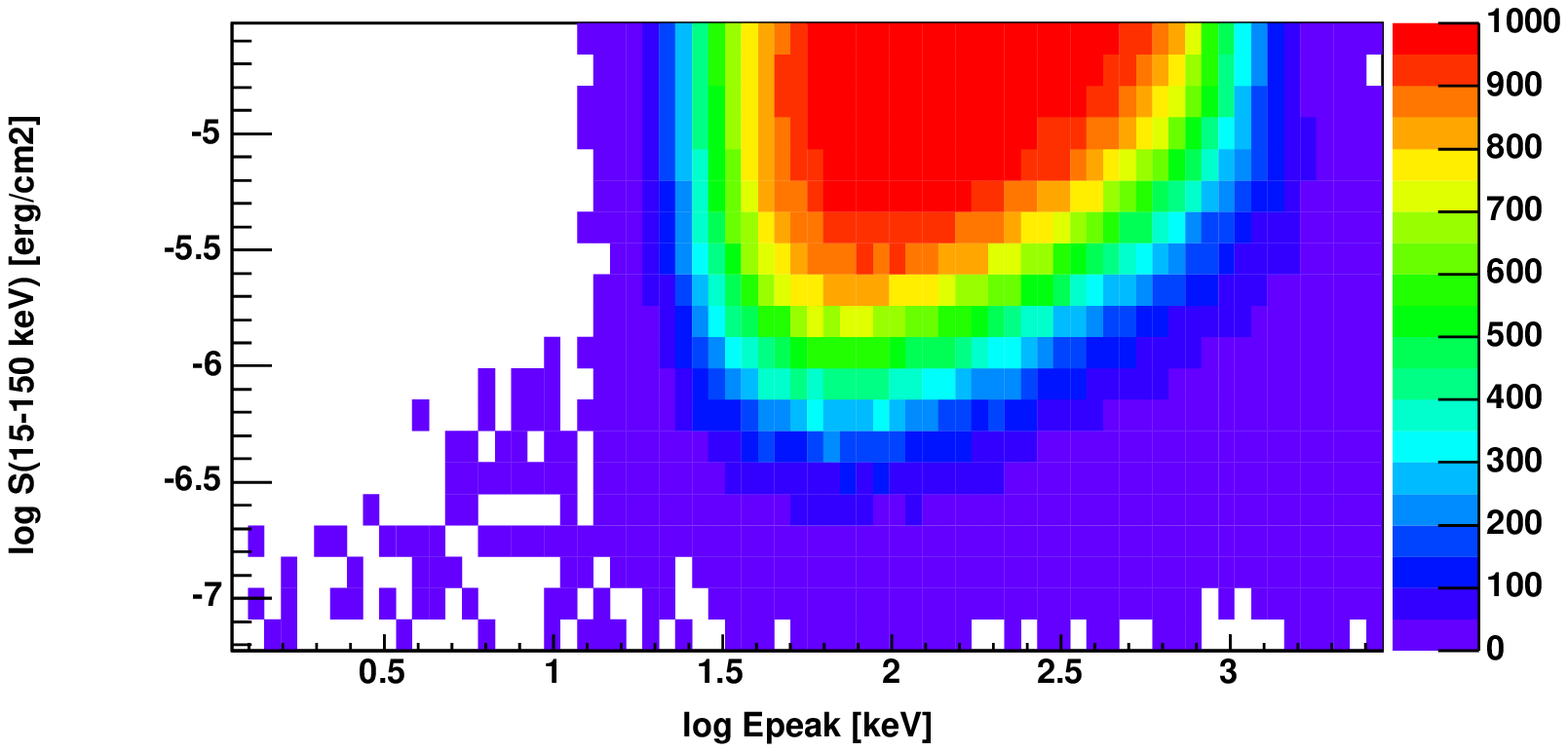}
\hspace{-1.0cm}
\includegraphics[scale=0.5]{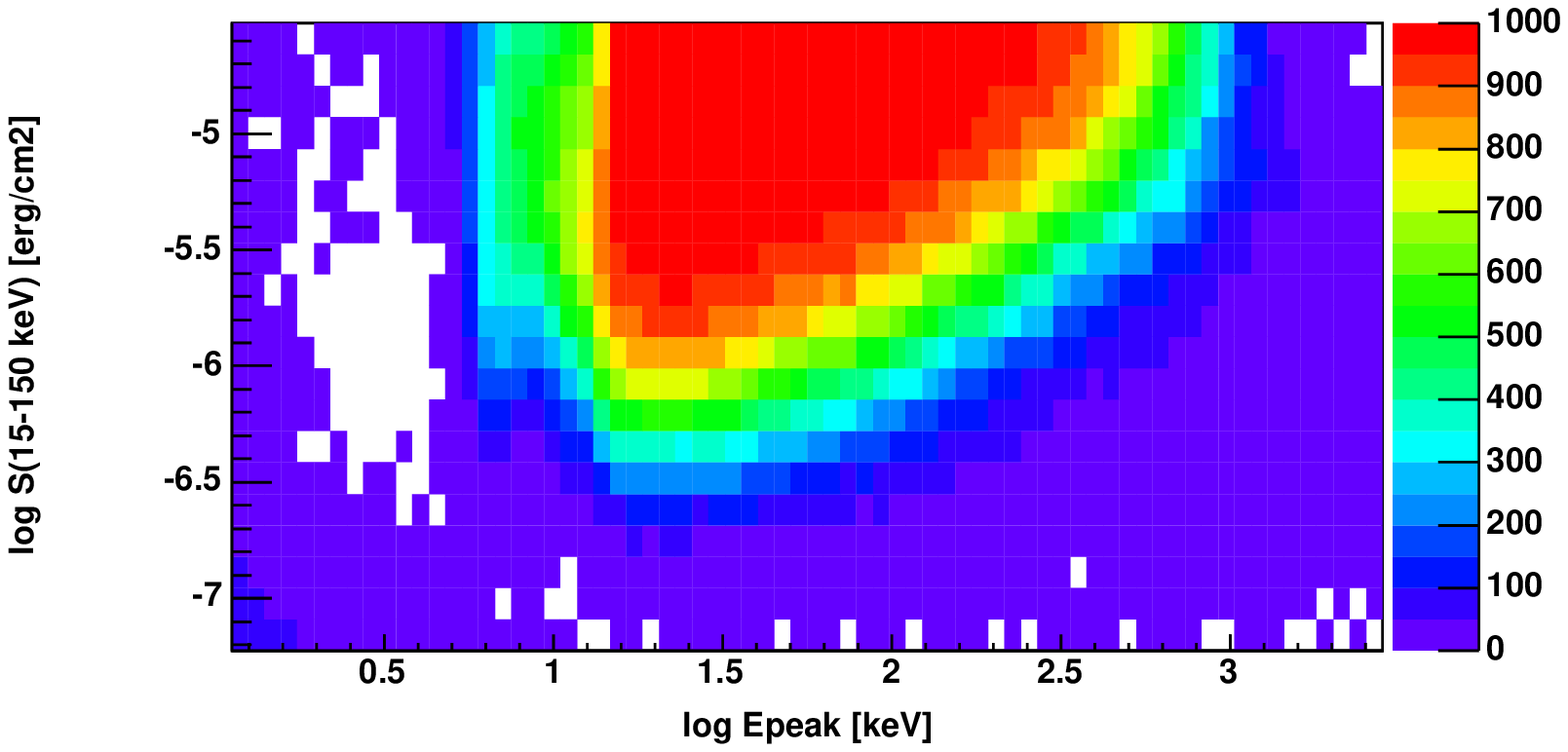}}
\centerline{
\includegraphics[scale=0.5]{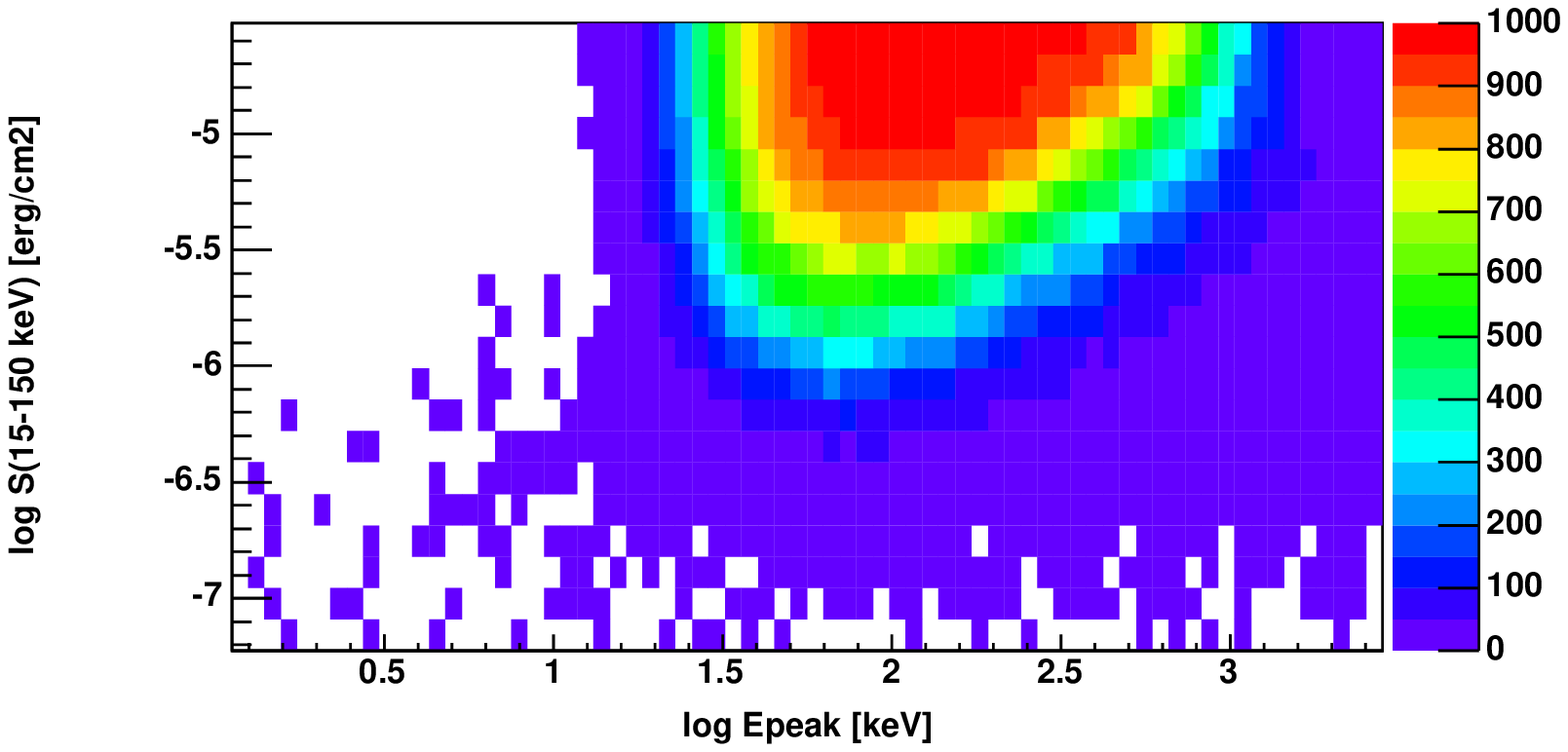}
\hspace{-1.0cm}
\includegraphics[scale=0.5]{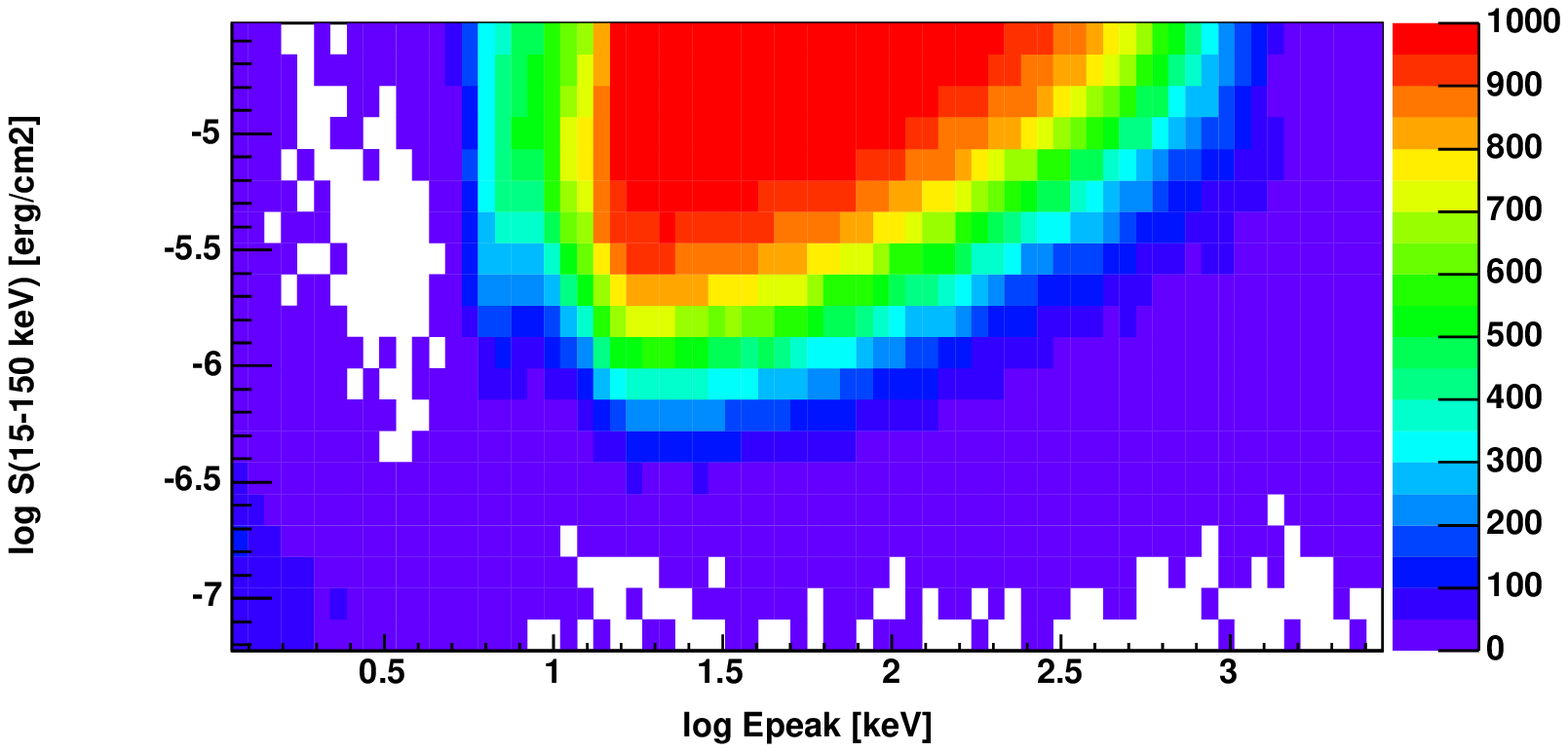}}
\centerline{
\includegraphics[scale=0.5]{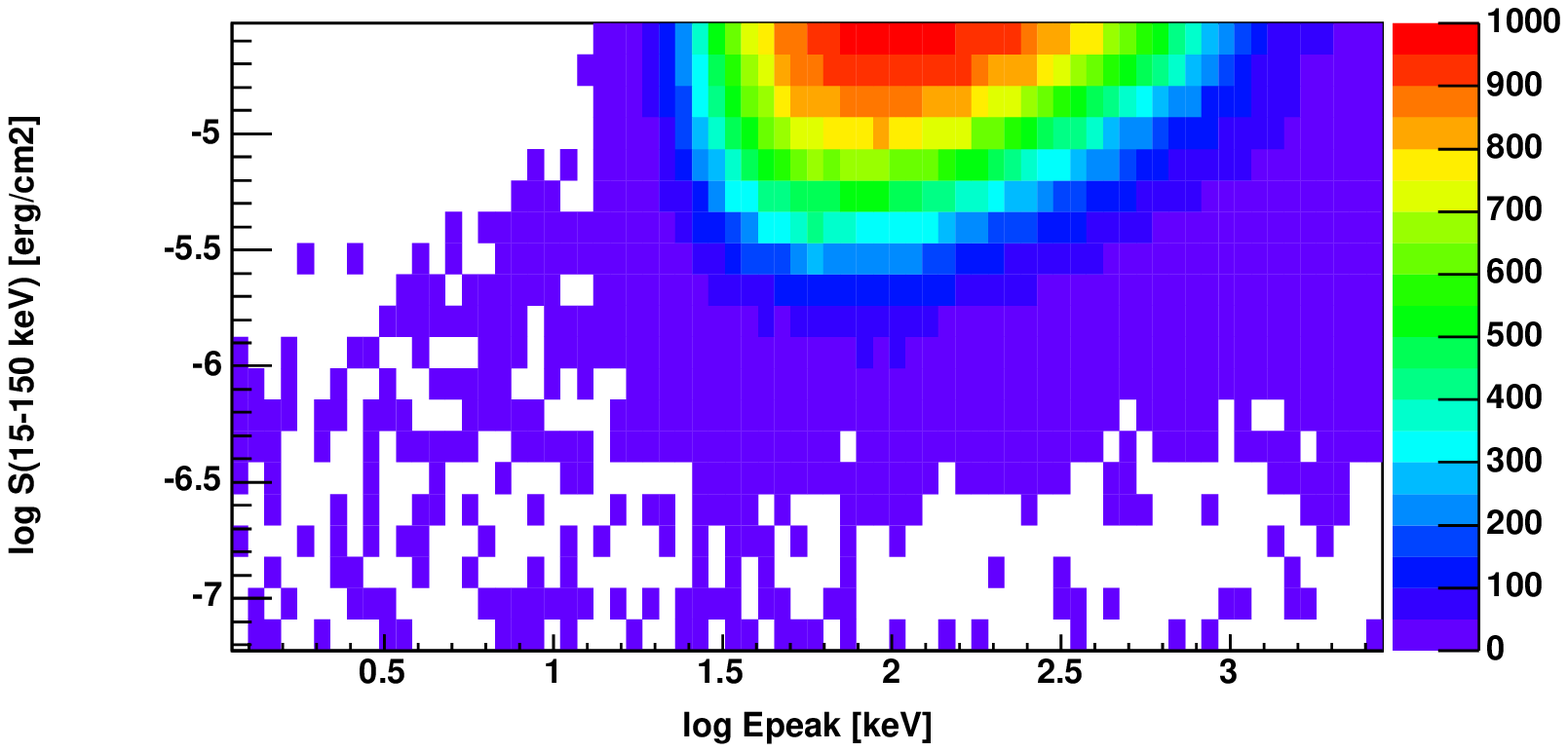}
\hspace{-1.0cm}
\includegraphics[scale=0.5]{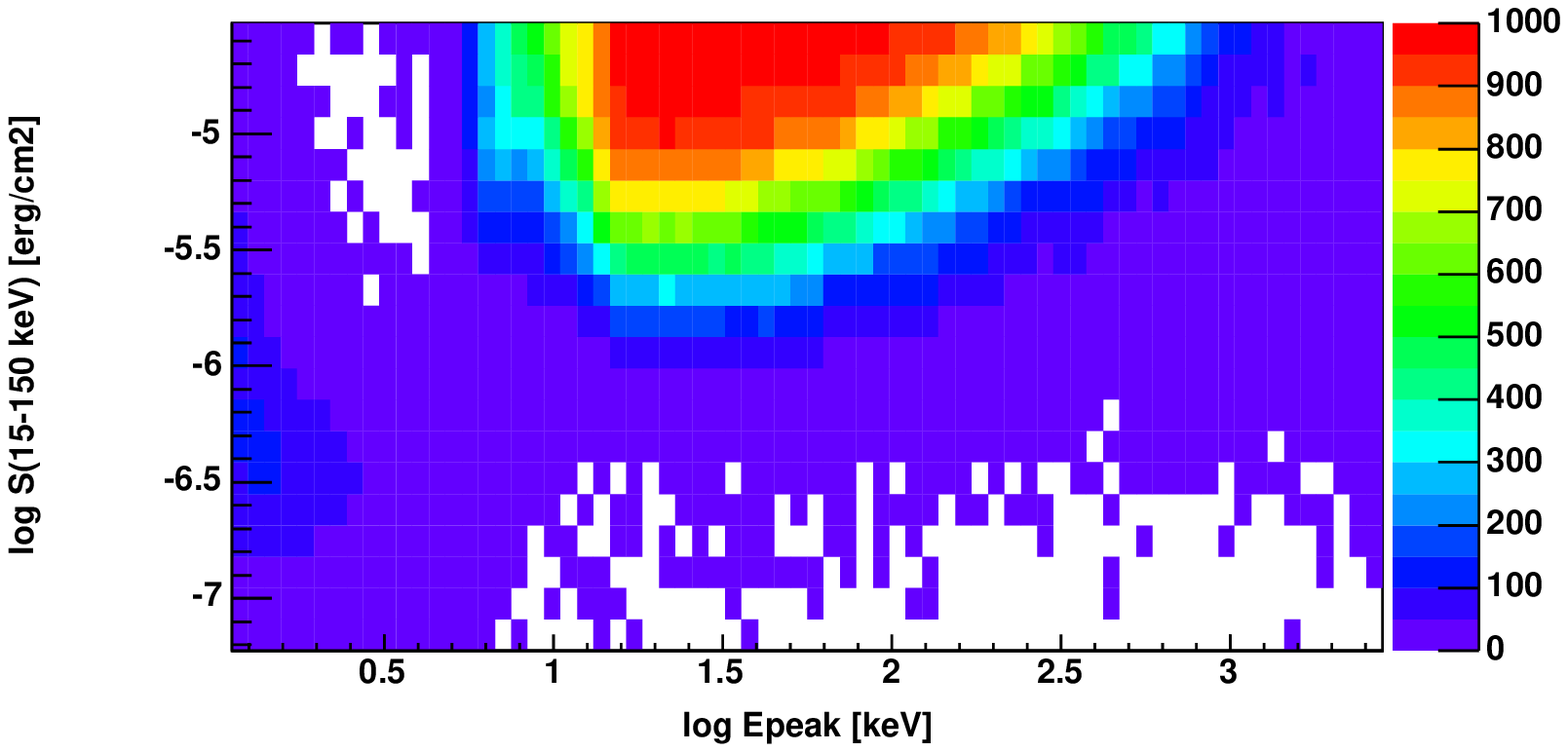}}
\caption{The contour maps showing, as a function of the simulated $\ep$ and the 
energy flux in the 15-150 keV band, the number of simulated spectra which have 
$\Delta\chi^{2}$ $>$ 6 (left row for the Band function: $\Delta\chi^{2} \equiv \chi^{2}_{\rm PL} - \chi^{2}_{\rm Band}$; 
right row for a CPL model: $\Delta\chi^{2} \equiv \chi^{2}_{\rm PL} - \chi^{2}_{\rm CPL}$).  The incident angles of 
the simulations are 0$^{\circ}$, 15$^{\circ}$, 30$^{\circ}$ and 50$^{\circ}$ from top to bottom.}
\label{fig:sim_ep_fluence}
\end{figure}

\begin{figure}
\centerline{
\includegraphics[scale=0.6,angle=-90]{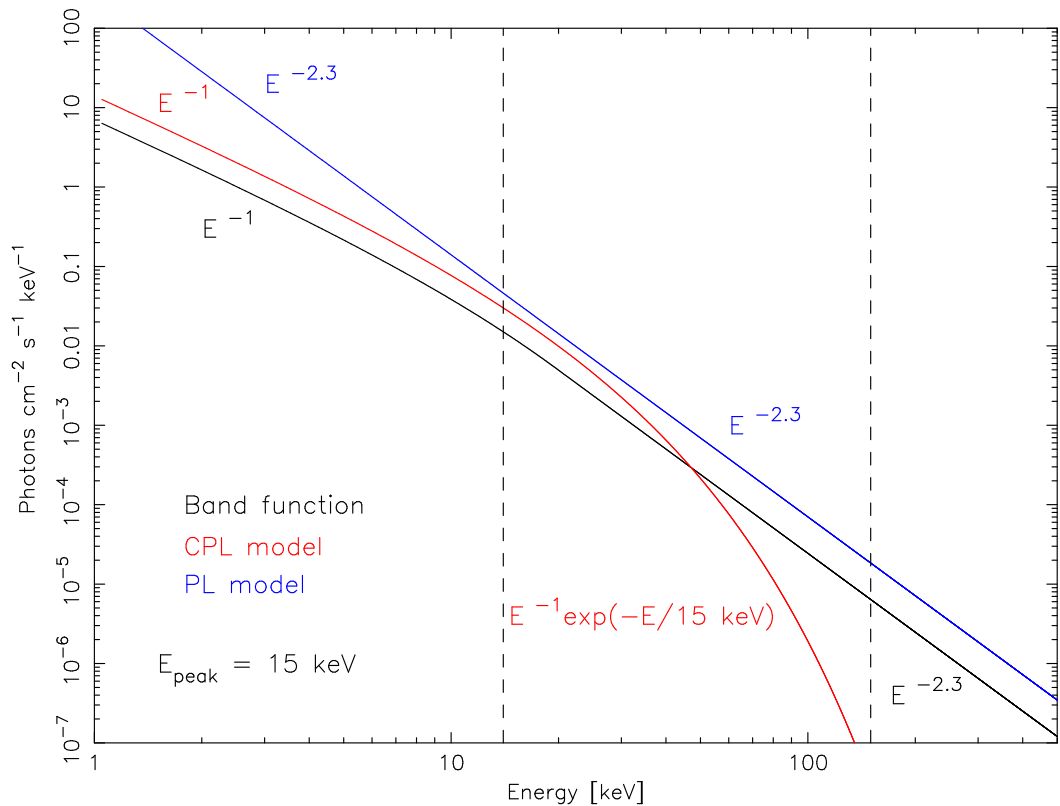}}
\caption{The schematic drawing of the photon spectra of the Band 
 function (black) and a CPL model (red) with $\ep$ of 15 keV.  The low 
energy photon index is $-1$ for both models.  The high energy photon
 index of the Band function is $-2.3$.  A PL model with a photon index 
of $-2.3$ is also overlaid in the plot (blue).  The vertical dotted lines 
are the BAT observed energy band of 15-150 keV.} 
\label{fig:exp_band_cpl_lowep}
\end{figure}

Figure \ref{fig:sim_ep_fluence} shows the numbers of the simulated spectra 
which have $\Delta \chi^{2}$ (
$\Delta \chi^{2}$ $\equiv \chi^{2}_{\rm PL} - \chi^{2}_{\rm Band}$
for the Band function or 
$\Delta \chi^{2}$ $\equiv \chi^{2}_{\rm PL} - \chi^{2}_{\rm CPL}$ for 
a CPL model) greater than 6\footnote{This is a current criterion used in
the BAT team for reporting the spectral parameters based on a CPL fit in
the BAT refined circular of the Gamma-ray Burst Coordinates Network.} as
a function of the $\ep$ and the energy flux in the 15-150 keV band.  
This $\Delta \chi^{2}$ $>$ 6 corresponds to $>$2.4 $\sigma$ confidence. 
The figures in the left and right row show 
the results based on the Band function and a CPL model, respectively.  
We note the distinct differences in the shapes of the confidence contours, 
especially 
at low $\ep$, between the Band function and a CPL model as an input spectrum.  
The results show that if a CPL model is indeed a true spectral shape, 
BAT can measure $\ep$ at the lower boundary of its energy range 
($\sim$ 15 keV) with a very high significance.  On the other hand, a low 
$\ep$ measurement would be very challenging if the Band function is the 
true spectral shape.  Figure \ref{fig:exp_band_cpl_lowep} explains the 
reason for these differences.  The figure shows the calculated photon spectra 
in a CPL model and the Band function for $\ep$ = 15 keV.  In a CPL model, 
the spectrum can not be fit with a PL model because of the curved shape 
(exponential component) in the BAT observed energy band.  Therefore, we
would expect a significant improvement in $\chi^{2}$ with a CPL 
fit over a PL fit.  However, in the Band function, due to the extra 
power-law component (high energy power-law component) in the formula,  
the spectrum at the BAT observed energy band would be just a simple 
power-law with a high energy photon index.  This is the reason why we 
see a difference in the confidence contours based on the assumed 
spectral models.  The results also show that the $\ep$ 
measurement becomes difficult for BAT when $\ep$ is below 30 keV or
above 100 keV in the Band function shape.  In the CPL shape, $\ep$ can be 
determined even at $\sim$ 15 keV.  

\begin{figure}
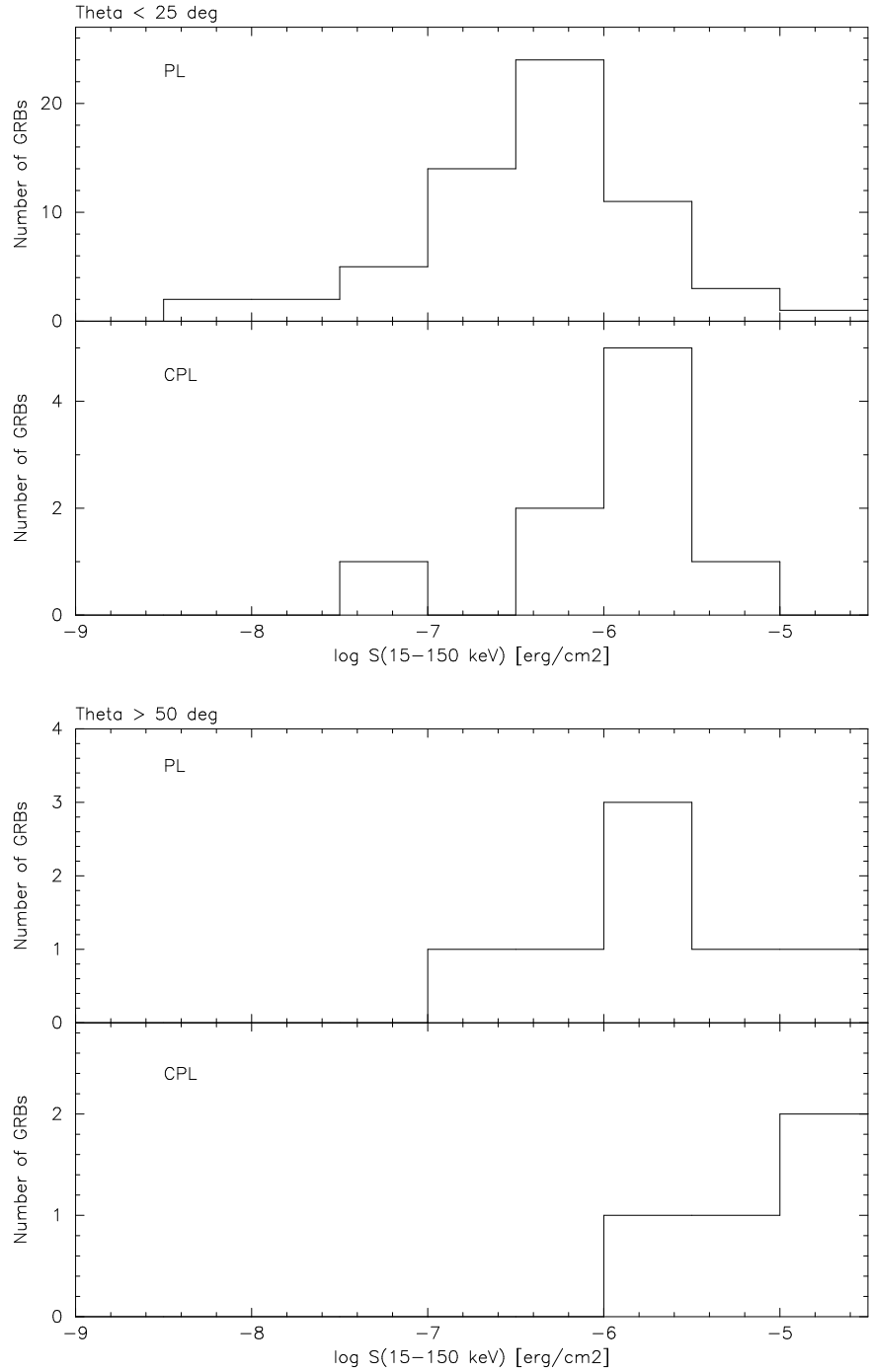

\centerline{
\includegraphics[scale=0.5,angle=-90]{f4a.eps}}
\vspace{0.5cm}
\centerline{
\includegraphics[scale=0.5,angle=-90]{f4b.eps}}
\caption{The number of GRBs acceptably fit by a PL model (top panel) and by a CPL 
model (bottom panel) as a function of the fluence in the 15-150 keV band.  The samples 
of the incident angles of bursts less than 25 degrees (top) and larger than 
50 degrees (bottom).}
\label{fig:hist_pl_cpl}
\end{figure}

Figure \ref{fig:hist_pl_cpl} shows the number of the BAT GRBs which can be 
acceptably fit by a PL model and by a CPL model as a function of the 
15-150 keV fluence.  The data are from the BAT1 catalog.  
In the case of an incident angle less than 25 degrees, a CPL 
model becomes an acceptable fit for fluence $>$10$^{-6}$ ergs cm$^{-2}$.  
However, a PL model still be acceptable fit if $\ep$ is located above or below 
the BAT energy range.  On the other hand, the fluence must typically be greater 
than $3 \times 10^{-6}$ ergs cm$^{-2}$ in the case of an incident 
angle greater than 50 degrees.
These threshold fluences required to measure $\ep$ in the BAT data 
correspond to the $\sim$50\% confidence contour (green) in our 
simulation results of Figure \ref{fig:sim_ep_fluence}.  

\begin{figure}
\centerline{
\includegraphics[scale=0.5]{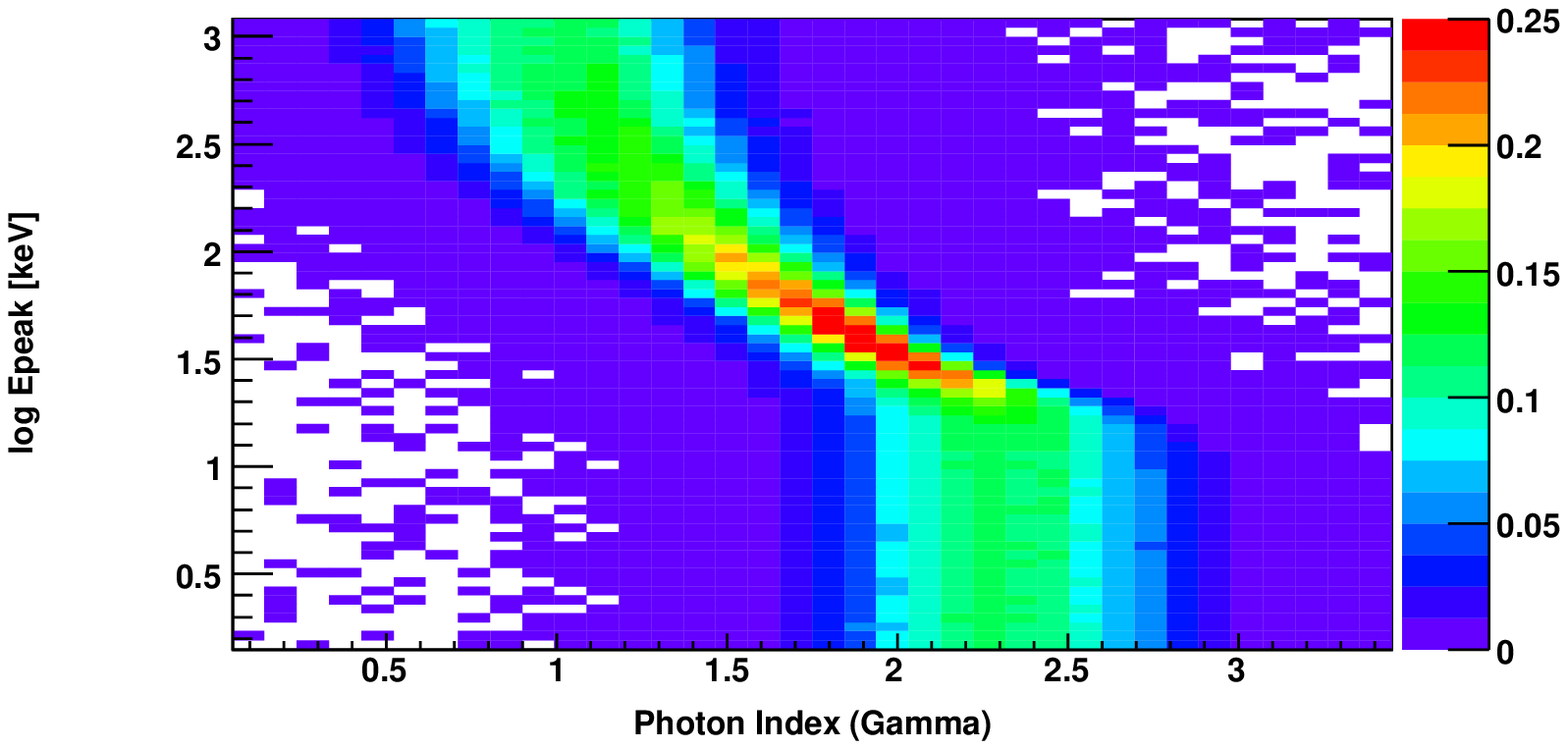}
\hspace{-1.0cm}
\includegraphics[scale=0.5]{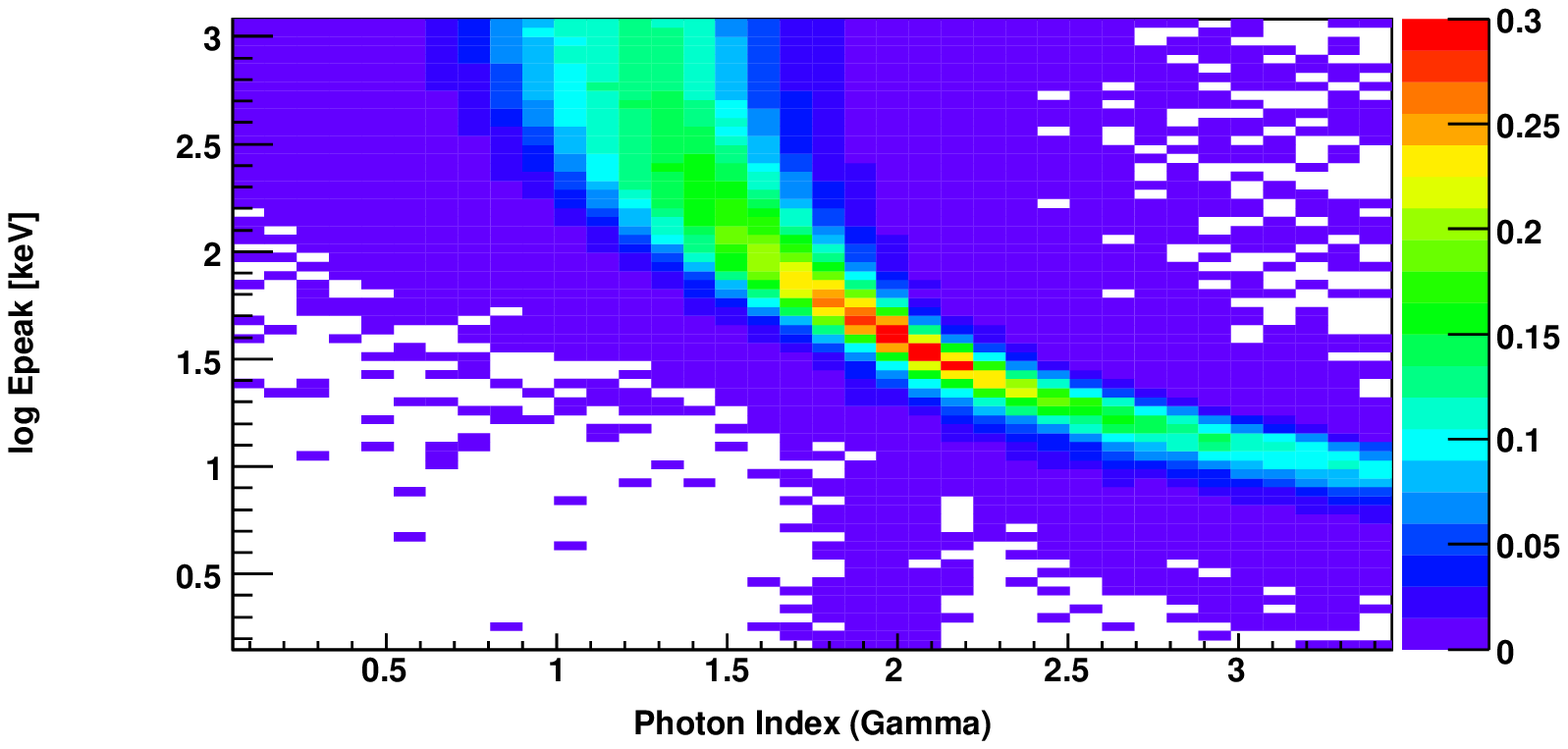}}
\centerline{
\includegraphics[scale=0.5]{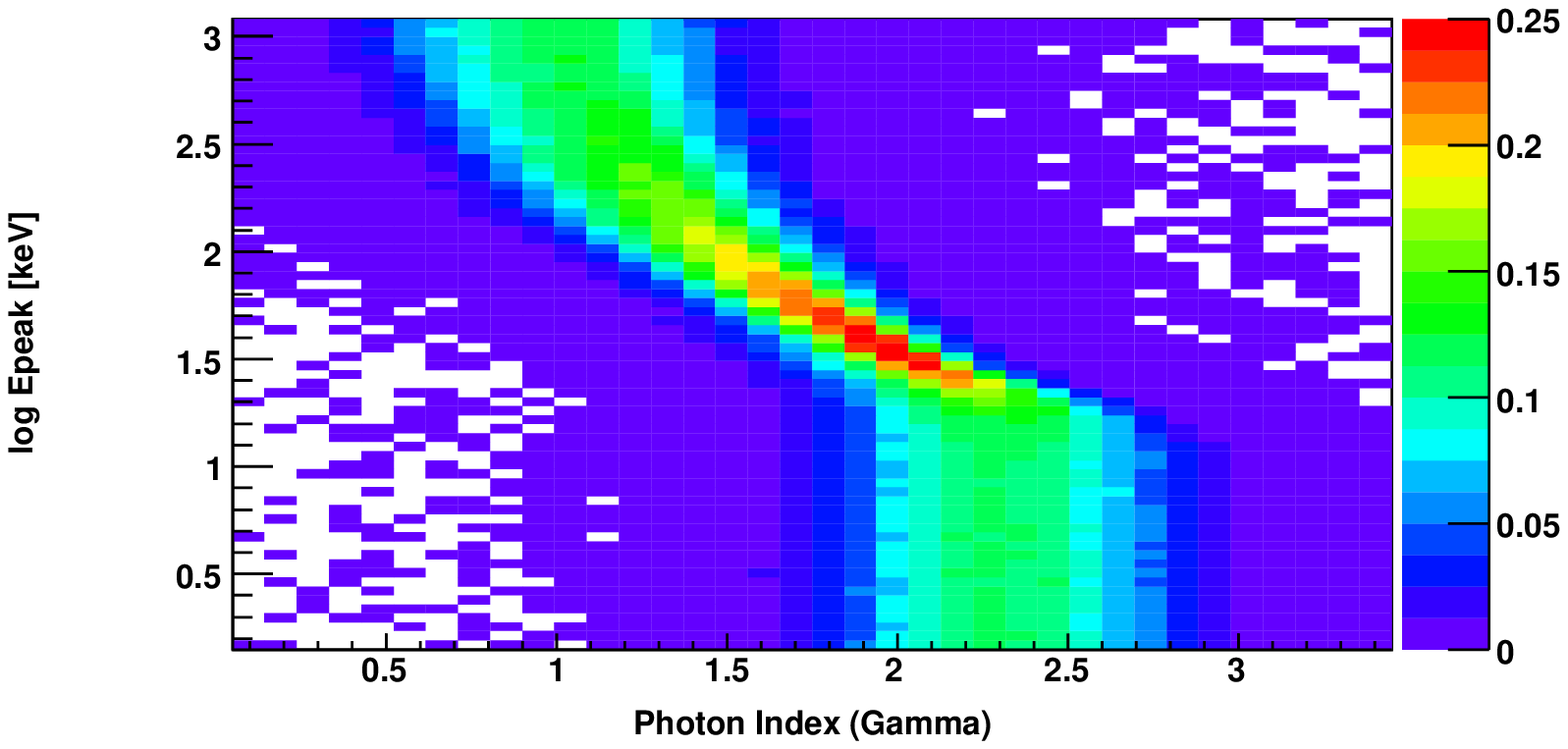}
\hspace{-1.0cm}
\includegraphics[scale=0.5]{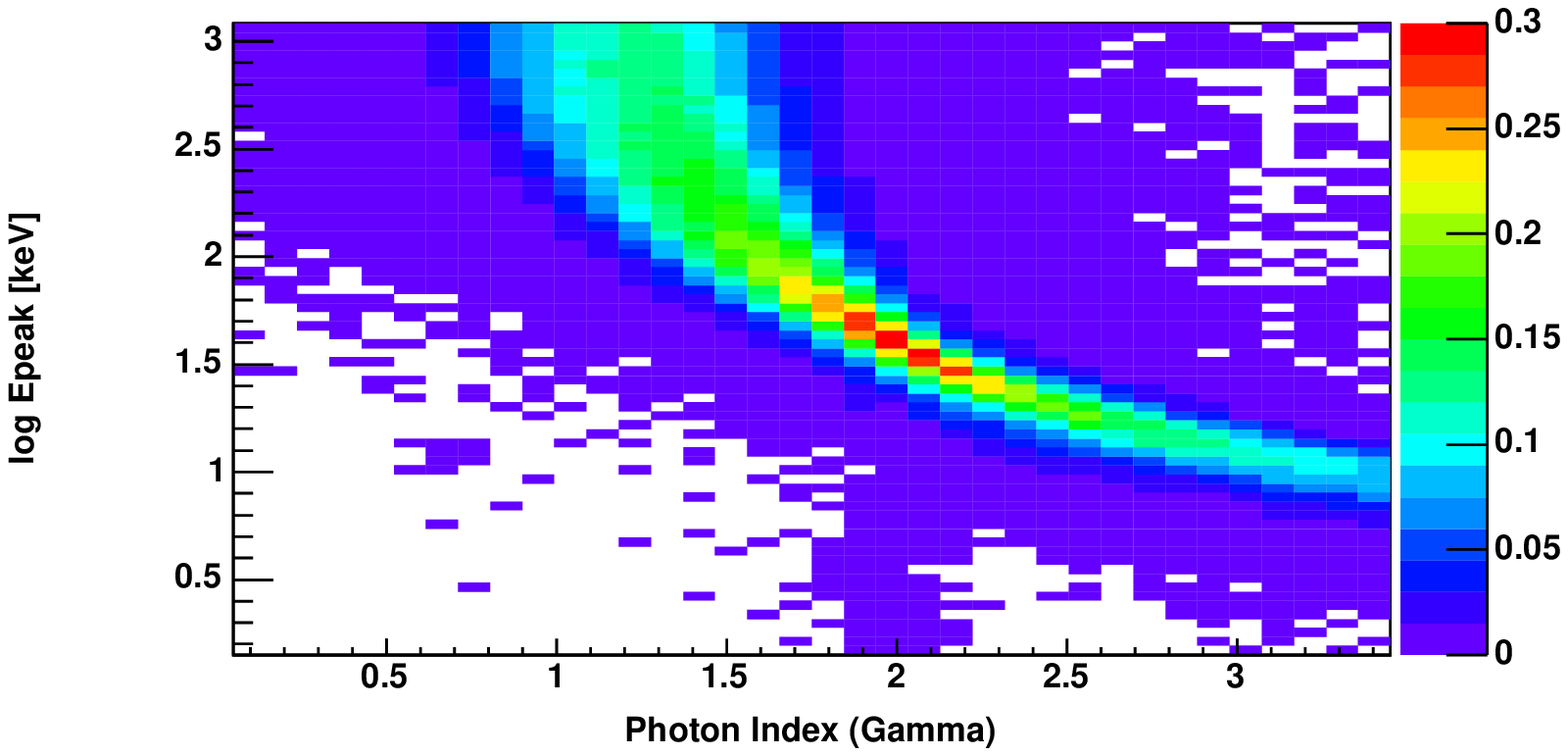}}
\centerline{
\includegraphics[scale=0.5]{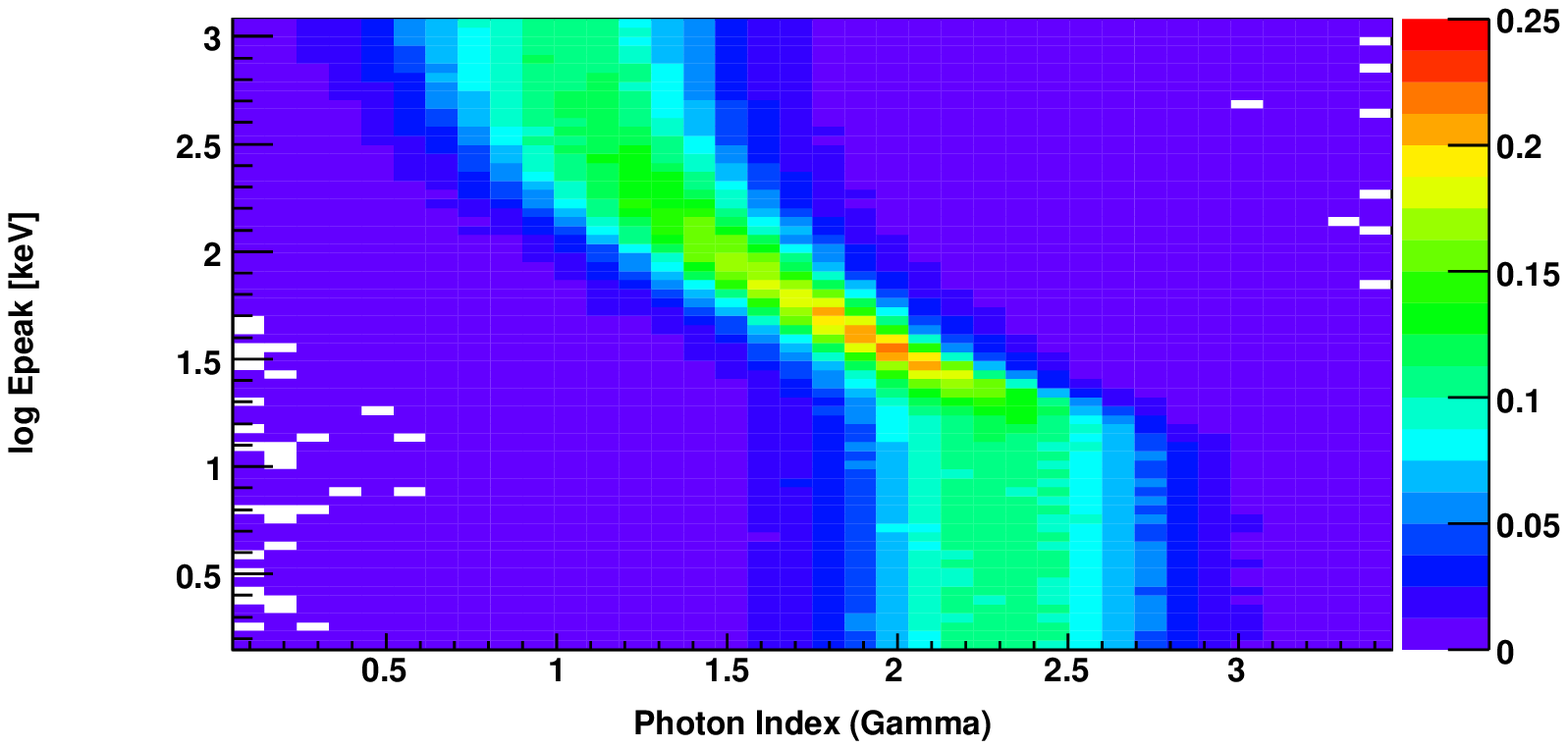}
\hspace{-1.0cm}
\includegraphics[scale=0.5]{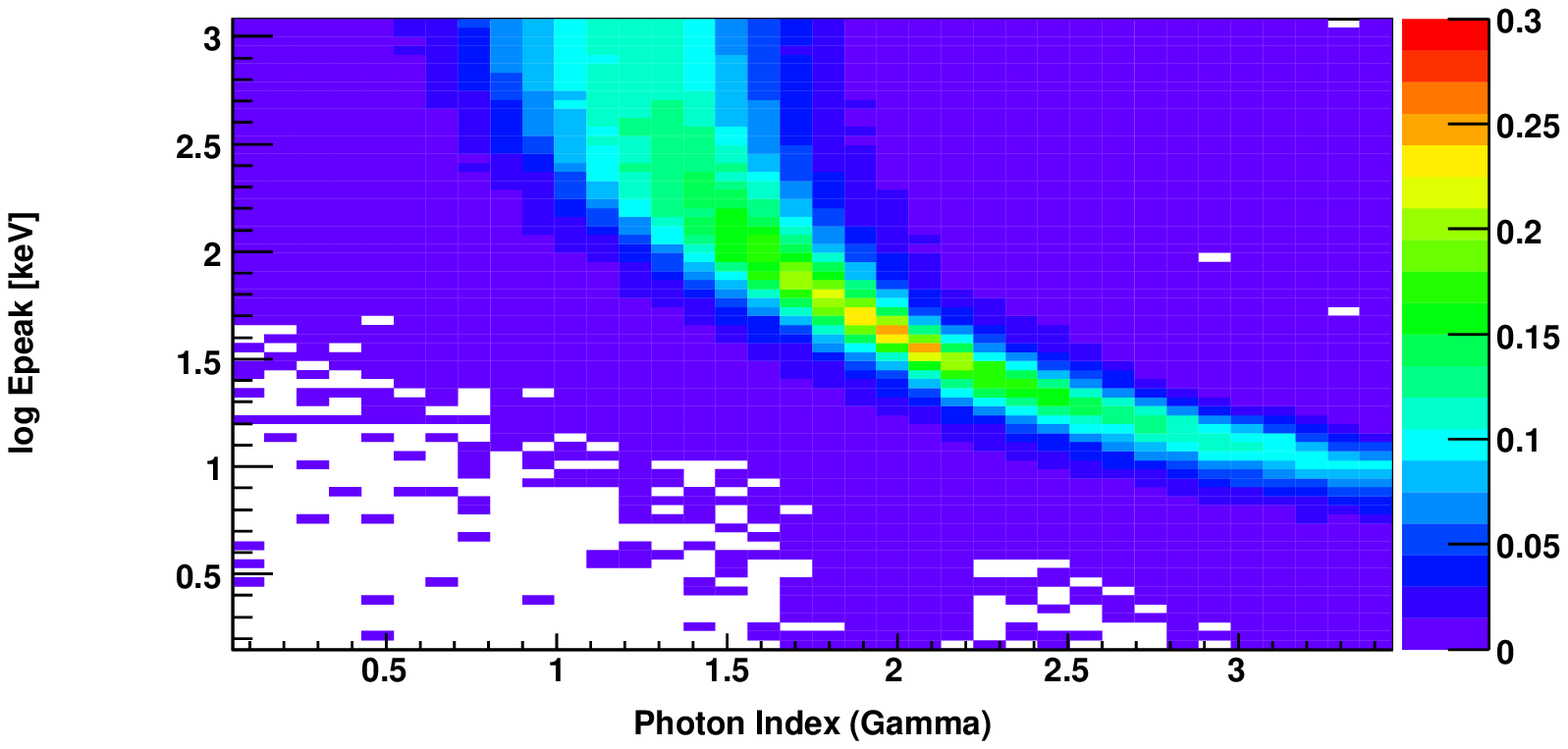}}
\centerline{
\includegraphics[scale=0.5]{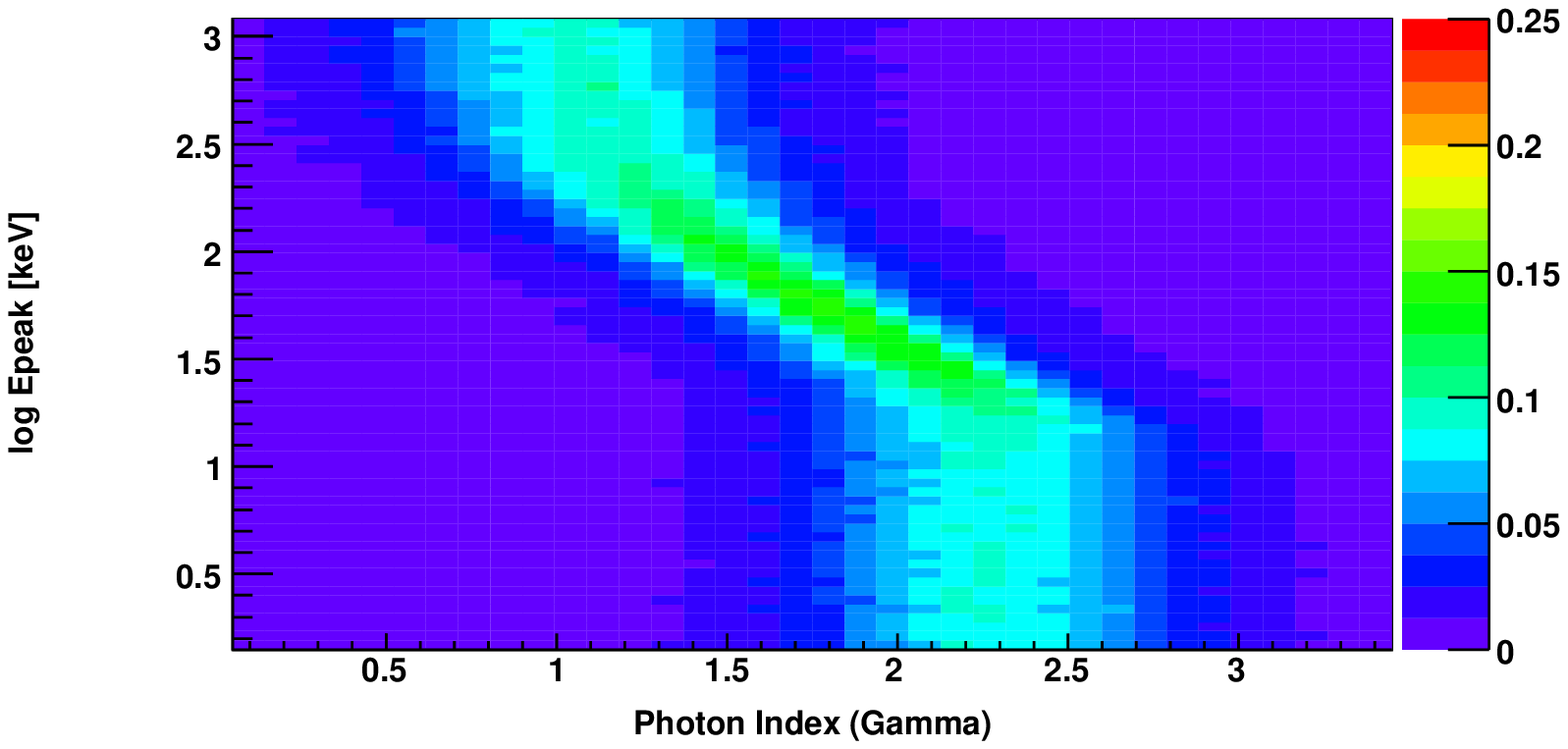}
\hspace{-1.0cm}
\includegraphics[scale=0.5]{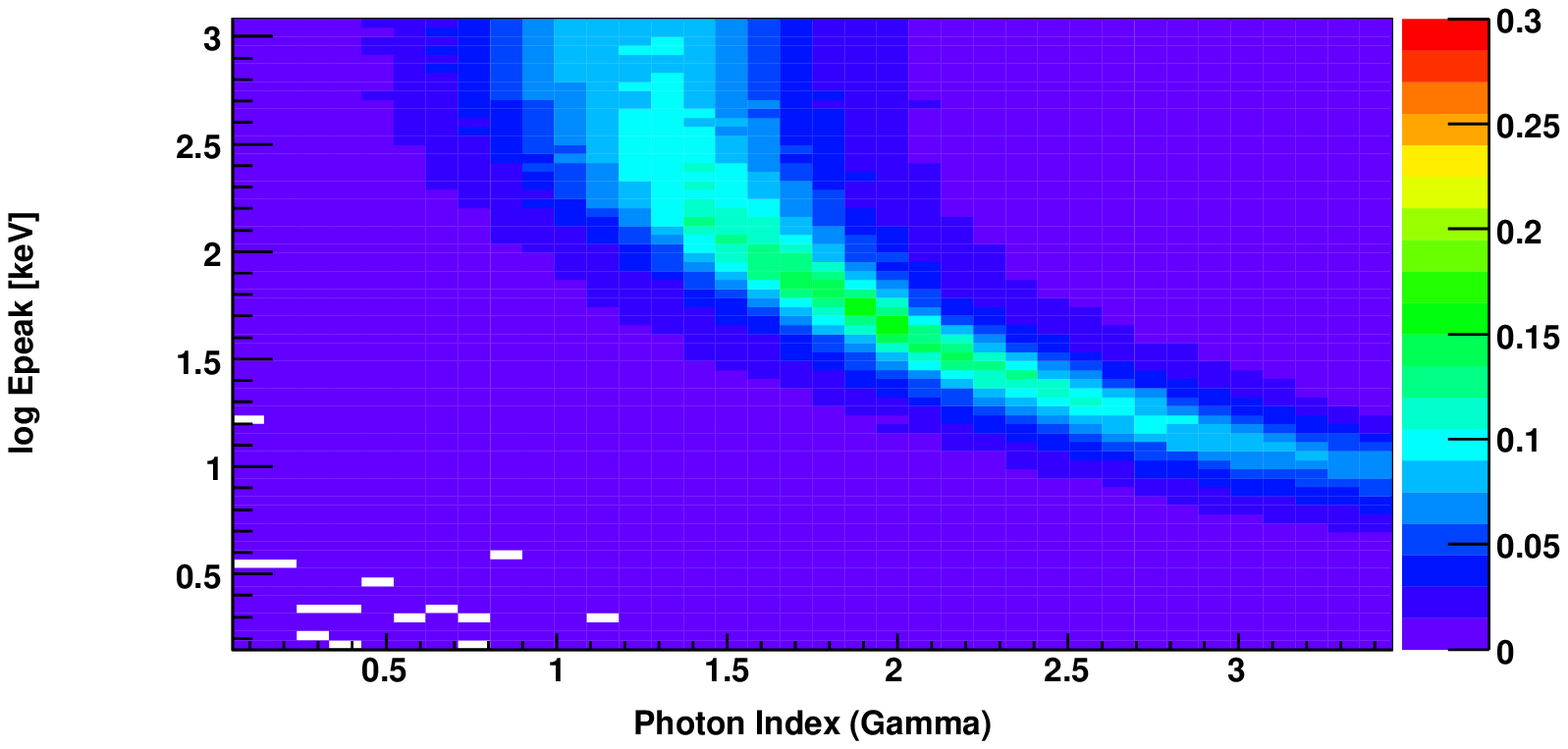}}
\caption{Contour maps showing the number of simulated spectra as a 
function of photon index and 
input $\ep$.  The left and right rows are the Band function and 
a CPL model, respectively.  The incident angles of the simulations 
are 0$^{\circ}$, 15$^{\circ}$, 30$^{\circ}$, and 50$^{\circ}$ 
from top to bottom.}
\label{fig:sim_phindex}
\end{figure}

Next, we made histograms of $\ep$ for each photon index on a 0.1 grid 
from 0 to 3.5 using the 
range of fluences corresponding to 
the 1-$\sigma$ interval of the BAT observed fluence distribution 
in the BAT1 catalog.  The 1-$\sigma$ fluence 
interval corresponds to the 
range from $3.4 \times 10^{-7}$ ergs cm$^{-2}$ to $5.4 \times 10^{-6}$ ergs cm$^{-2}$.  
This selection 
of the fluence range allows us to reduce the systematic effect of the inclusion 
of unrealistically bright or dim simulations.  
Furthermore, since we are interested in estimating the $\ep$ for the bursts which 
do not show a significant improvement in $\chi^{2}$ by a CPL fit over a PL fit, 
we also only selected the simulated spectra with $\Delta \chi^{2}$ 
$=$ $\chi^{2}_{\rm PL} - \chi^{2}_{\rm CPL}$ $< 6$.  Because the numbers of simulated 
spectra are different for each $\ep$ grid due to these selections, we normalized the 
number of simulated spectra in each $\Gamma$-$\ep$ grid by the total number of spectra 
in each $\ep$ grid.
Figure \ref{fig:sim_phindex} shows the 
contour map of the photon index ($\Gamma$) and $\log \ep$ for the Band function (left)
and a CPL (right) model.  There is a 
correlation between $\Gamma$ and $\log \ep$ in the range from 1.3 to 2.3 of 
$\Gamma$ for the Band function.  The correlation continues to $\Gamma = 3.0$ in the 
case of a CPL model for the same reason as we demonstrated in Figure \ref{fig:exp_band_cpl_lowep}.  
It might be interesting to note that a 
very steep photon index such as $\Gamma$ $\sim$ 3 is not possible to achieve 
if the source spectrum is the Band function.  In this case, the source 
spectrum might be much closer to a CPL shape.  One important conclusion is that 
the correlation between $\log \ep$ and $\Gamma$ exists independent of the 
incident angle of the burst.  Therefore, this correlation, the $\ep$ - $\Gamma$ 
relation, can be used for all BAT long GRBs within the allowed $\Gamma$ range, 
although with larger uncertainty for GRBs at large incidence angle.  

\begin{figure}
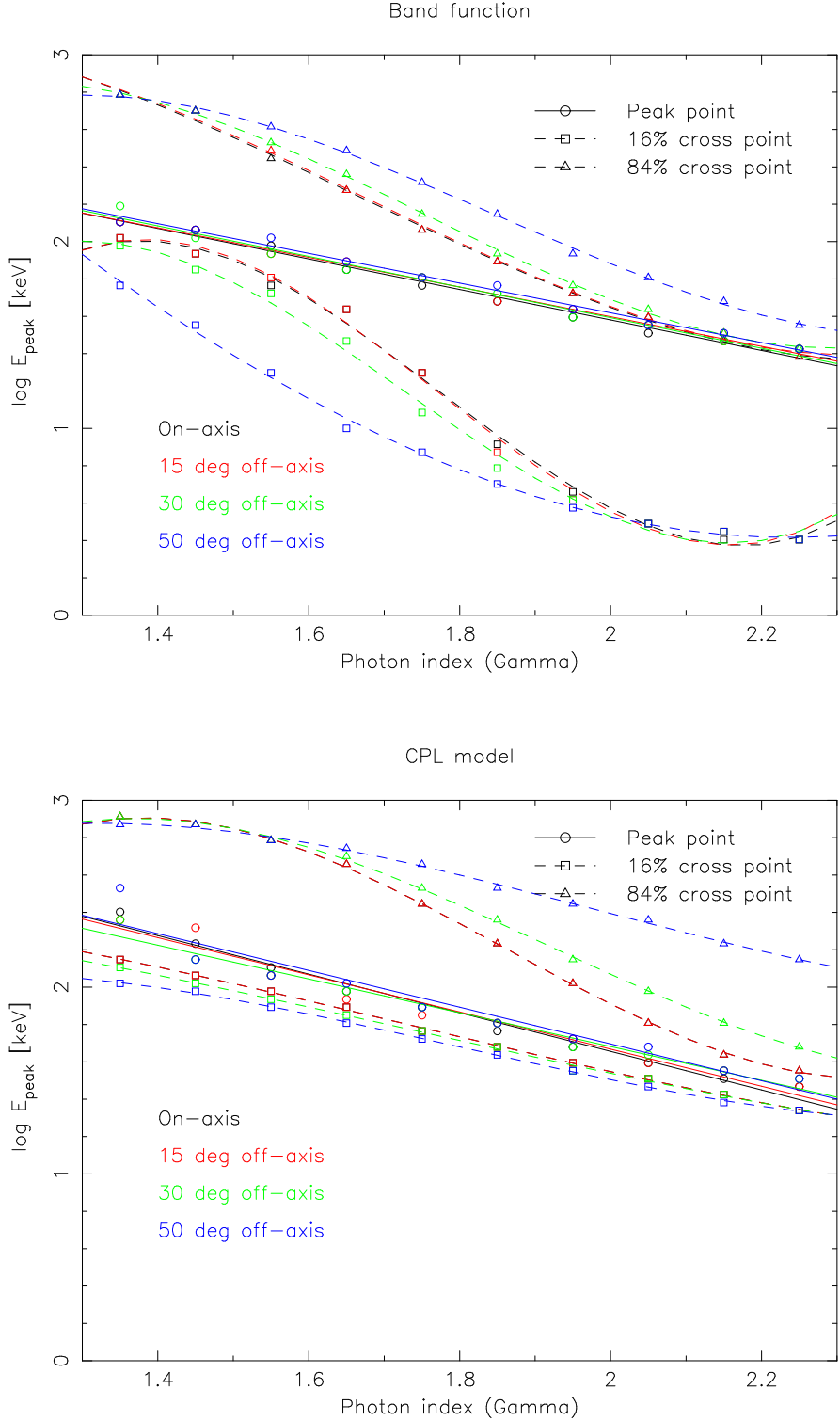

\centerline{
\includegraphics[scale=0.5,angle=-90]{f6a.eps}}
\vspace{1cm}
\centerline{
\includegraphics[scale=0.5,angle=-90]{f6b.eps}}
\caption{The best fit $\ep$ - $\Gamma$ relations (solid line) and the 
lower and higher 1-$\sigma$ confidence level of the relations (dashed
 lines) for the Band function (top) and a CPL model (bottom) with the
 data points (circles: $\ep$ at the peak of the histogram of
 $\Gamma$, squares: $\ep$ value of 16\% crossing point of the histogram
 of $\Gamma$, and triangles: $\ep$ value of 84\% crossing point of the histogram
 of $\Gamma$).  The black, red, green and blue show the cases of 
 incident angles 0$^{\circ}$, 15$^{\circ}$, 30$^{\circ}$ and
 50$^{\circ}$, respectively.}
\label{fig:ep_gamma_fit}
\end{figure}

\begin{figure}
\centerline{
\includegraphics[scale=0.6,angle=-90]{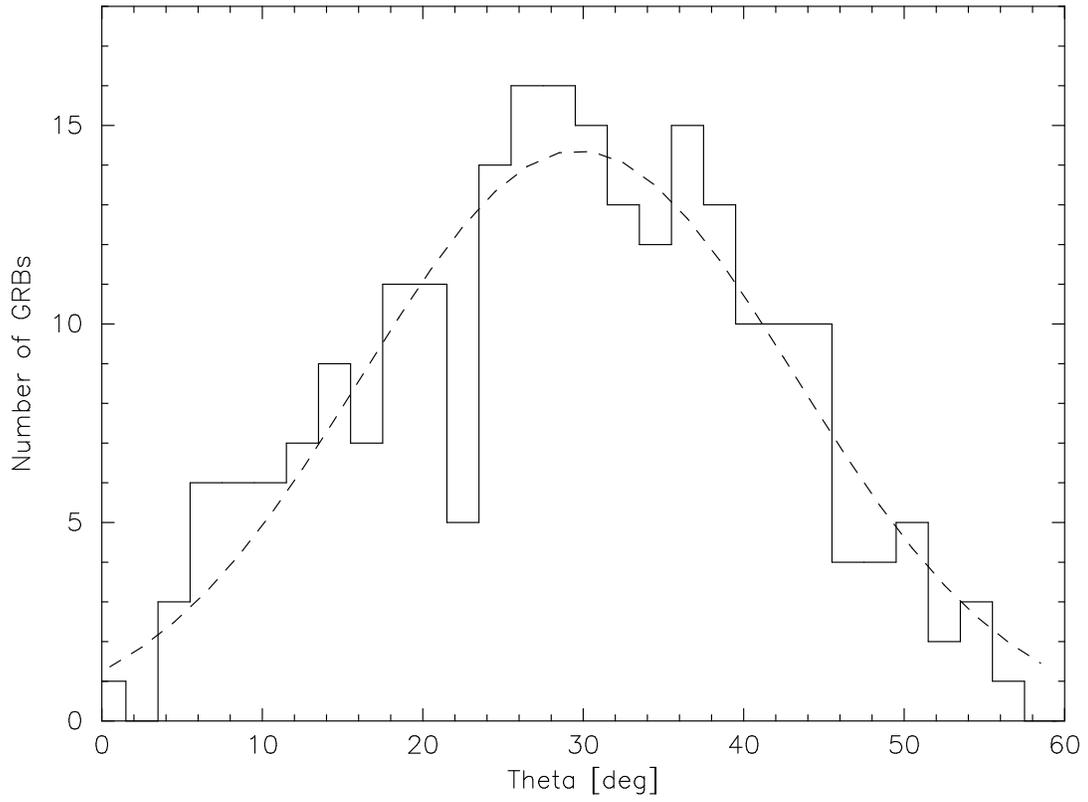}}
\caption{The incident angle ($\theta$) distribution of the BAT GRBs.
 The dotted line is the best fit gaussian model.}
\label{fig:theta}
\end{figure}

\begin{figure}
\centerline{
\includegraphics[scale=0.9]{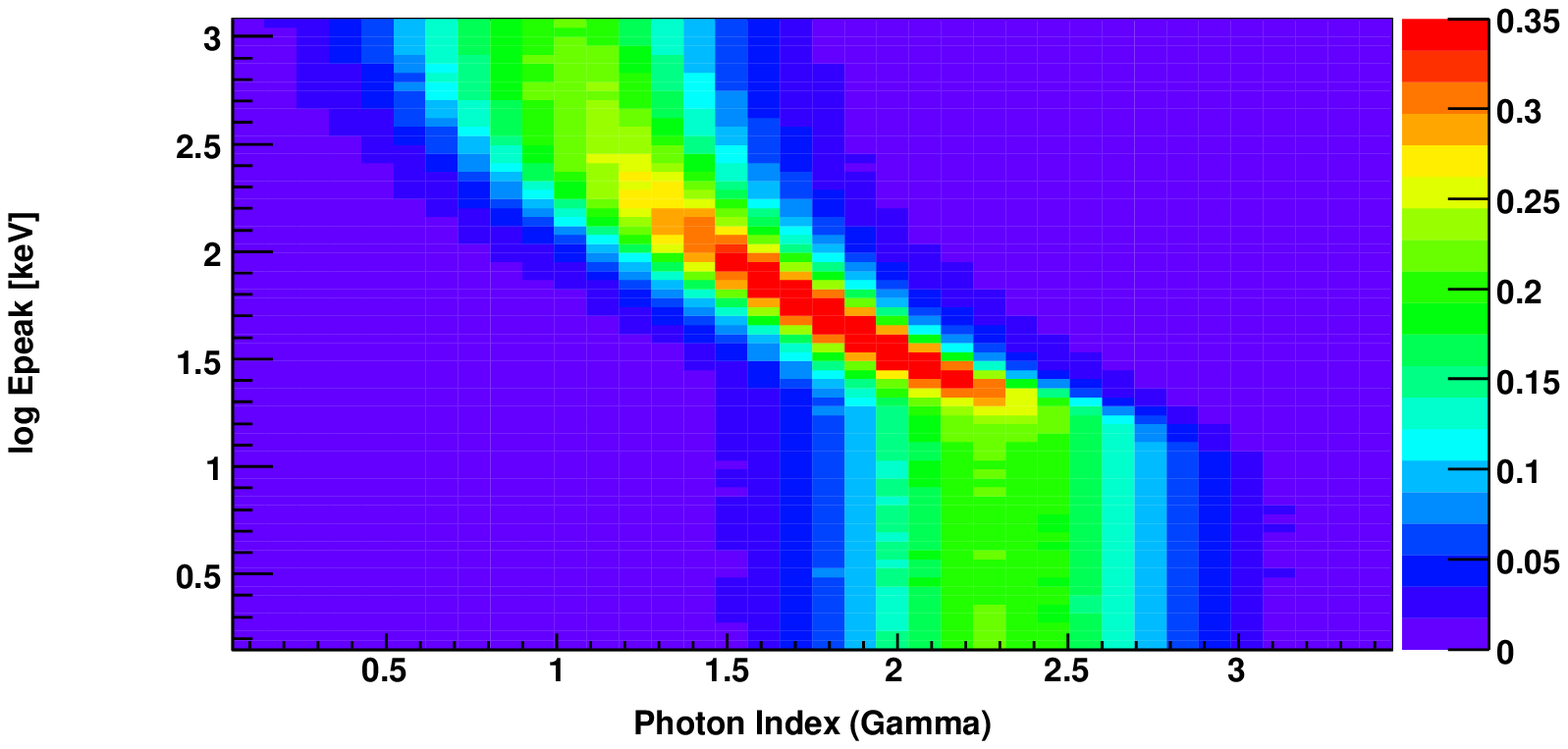}}
\vspace{1cm}
\centerline{
\includegraphics[scale=0.9]{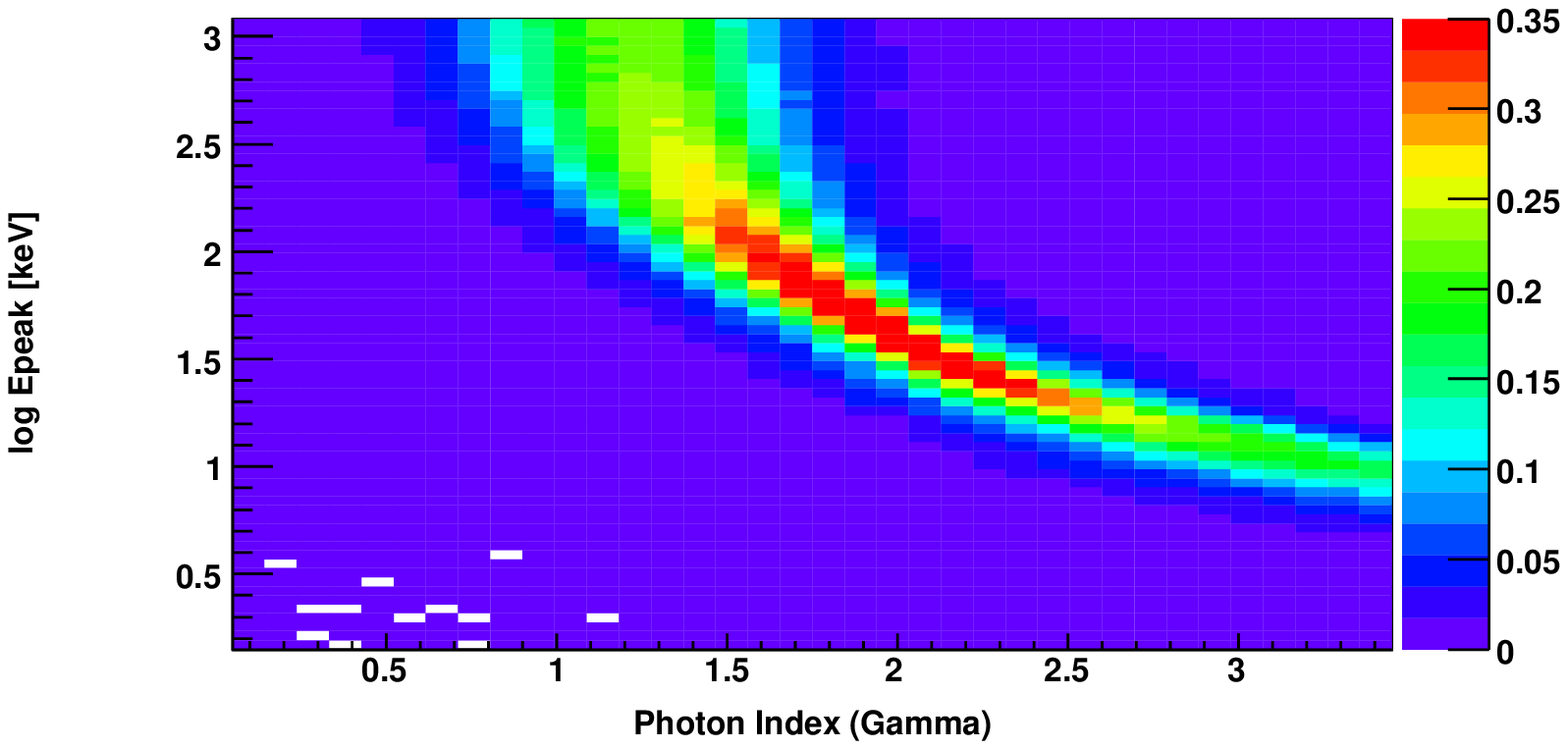}}
\caption{Contour maps showing the number of simulated spectra as a 
function of photon index and input $\ep$ after weighting the 
simulation results of 0$^{\circ}$, 15$^{\circ}$, 30$^{\circ}$, 
and 50$^{\circ}$ incident angles by the incident angle distribution 
of the BAT GRBs shown in Figure \ref{fig:theta} (top: the Band function and bottom: 
a CPL model).}
\label{fig:weighted_ep_gamma}
\end{figure}

We extracted the peak $\ep$ value from each histogram of $\Gamma$ 
and fit with a linear function using the range from $1.3 < \Gamma < 2.3$ 
for both the Band function and a CPL model.  
Although the correlation exists until $\Gamma = 3$ in a CPL case, 
we use the same $\Gamma$ range for the Band function and a CPL model 
to investigate the systematic difference based on the assumption of the 
source spectrum.  
The best fit $\ep$-$\Gamma$ relations are summarized in Table
\ref{tbl:ep_gamma_band} (Band function) and Table \ref{tbl:ep_gamma_cpl} (CPL).  
To estimate the 1-$\sigma$ uncertainty of the relation, we found 16\% and 84\% 
points of $\ep$ from each histogram of $\Gamma$ and fitted with a cubic function 
from $1.3 < \Gamma < 2.3$.  The best fit cubic functions of the lower and higher 
1-$\sigma$ confidence level are also summarized in Table
\ref{tbl:ep_gamma_band} and Table \ref{tbl:ep_gamma_cpl}. 
Figure \ref{fig:ep_gamma_fit} shows the best fit functions of 
the $\ep$ - $\Gamma$ relation and its 1-$\sigma$ confidence level with the data points 
used in the fittings.  
We note that the wide $\ep$ range in the simulations (in our case from 
1.4 keV to 1210 keV) is essential to derive the 1-$\sigma$ confidence level 
of the relation.  If the $\ep$ range in the simulations is not wide enough 
such as from 10 keV to 500 keV, we noticed that the confidence level will 
be underestimated by a factor of 2 for the upper limit at $\Gamma$ of 1.3 
and by a factor of 5 for the lower limit at $\Gamma$ of 2.0.  Due to the 
smoothly curved shapes of the Band function and the CPL model, the $\ep$ grids 
in the simulations have to be an order of magnitude wider than the energy 
range of the instrument, so that a curvature (or $\ep$) 
in the spectrum is completely outside the energy range of the instrument 
for the $\ep$ around the energy limits of the instrument.  
However, we also notice that the best fit $\ep$ - $\Gamma$ relation itself 
is less sensitive to the energy limits on the simulations.  
Although the confidence level is different between 
the Band function and a CPL model, the best fit linear function shows little 
difference between these two spectral models.  
We also calculated the relation weighting the
results at the incident angles of 0$^{\circ}$, 15$^{\circ}$,
30$^{\circ}$, and 50$^{\circ}$ by the distribution of 
the incident angle of the BAT GRBs (Figure \ref{fig:theta}).  Hereafter, 
we call this relation as the weighted $\ep$ - $\Gamma$ relation.  
The contour plots of the weighted $\ep$ - $\Gamma$ relation, the plot of the 
best fit functions, and the formula of the best fit functions are shown 
and summarized in Figure \ref{fig:weighted_ep_gamma}, Figure
\ref{fig:weighted_ep_gamma_fit}, and Table \ref{tbl:ep_gamma_band} and 
Table \ref{tbl:ep_gamma_cpl}, respectively.  

In the application of our $\ep$ - $\Gamma$ relation, we strongly encourage 
the reader to use the result based on the Band function as a prior.  The main 
reason for also performing the simulations of a CPL model as a prior is to see the 
systematic effect due to a prior assumption of the spectral model.  Our results 
are clearly demonstrating the effect of the assumed spectral model.  From the 
various measurements of the burst spectra by different instruments, the true 
burst spectrum is very likely to be the Band function at least for long GRBs.  
Therefore, the $\ep$ - $\Gamma$ relation based on the Band function as a prior 
is the most suitable relation to apply for the BAT long GRBs.  

\section{Comparison to other $\ep$ Measurements}

To investigate the validity of our simulation study, we used the
spectral parameters on the BAT1 catalog \citep{sakamoto2007}.  
Table \ref{hyo:spec_para} shows the spectral parameters of 31 long GRBs (T$_{90}$ $>$ 
2 seconds) having $\Delta \chi^{2}$ of greater than 6 in a CPL model over a PL model fit.   
Figure \ref{fig:comp_epeak_gamma_obs} shows $\ep$ energy in a CPL model
and the photon index, $\Gamma$, in a PL model for the BAT GRBs overlaid with the 
weighted $\ep$ - $\Gamma$ relation.  We also plot $\Gamma$ derived 
from the BAT data and $\ep$ reported by Konus-Wind or $HETE$-2 in the Gamma ray 
bursts Coordinates Network (GCN) listed in table \ref{hyo:spec_para_kw_hete2}.  
As seen in the figure, 
the 1-$\sigma$ confidence level of the $\ep$ - $\Gamma$ relation based on the simulation study 
is consistent with the 90\% confidence level of $\Gamma$ observed by the BAT 
and $\ep$ observed by the GRB instruments.  
However, we want to caution about using our $\ep$ - $\Gamma$
relation for estimating $\ep$.  Our estimator is based on prior
assumptions of the low-energy and/or the high-energy photon index 
measured by the BATSE.  Therefore, $\ep$ based on our $\ep$ - $\Gamma$ 
relation only provides a likelihood of the $\ep$ value not the ``measurement.''

The calculation of the bolometric flux or fluence is another challenge 
when using the 
BAT data alone.  However, since the low energy photon index $\alpha$ and the high energy 
photon index $\beta$ of the Band function are quite stable parameters even if $\ep$ varies from 
a few keV to a few MeV \citep[e.g.,][]{sakamoto2005,kaneko2006}, one could 
estimate the bolometric flux or fluence assuming the best fit $\alpha$ and $\beta$ 
from the BATSE time-averaged spectral analysis of \citet{kaneko2006}, and applying 
the best fit $\ep$ derived from our $\ep$ - $\Gamma$ relation.  
To get the normalization for the Band function spectrum, one would scale the Band 
spectrum so that the flux in the BAT energy range matches to the BAT measured flux.  
One can also estimate 
the error of the flux or fluence by propagating the errors of $\alpha$, $\beta$, $\ep$, 
and the normalization, however this estimate will not be strictly correct because 
the parameters of the Band function are correlated.  Finally, we caution against relying 
too heavily on this derived bolometric flux or fluence, since the method described uses 
averaged $\alpha$ and $\beta$ from a different burst population and an estimated, 
rather than measured $\ep$.

\section{Discussion}

According to \citet{sakamoto2005}, an equal number of 
X-ray flashes (XRF), X-ray-rich GRBs (XRR), and GRBs are 
reported in the HETE-2 GRB sample.  Their classification of GRBs based on the fluence 
ratio between the 2-30 keV and 30-400 keV bands is almost the equivalent of classifying 
GRBs by $\ep$.  The boundaries of $\ep$ between an XRF and 
an XRR, and an XRR and a GRB are around 30 keV and 100 keV.  
When we use the weighted $\ep$ - $\Gamma$ relation for the Band function to calculate 
the corresponding $\Gamma$ for each $\ep$, $\Gamma$ is $\sim$ 2.2 and 
$\sim$ 1.5 for $\ep$ of 30 keV and 100 keV, respectively.  Applying these $\Gamma$ 
criteria to a sample of 206 BAT 
GRBs, excluding short GRBs (T$_{90}$ $<$ 2 seconds) and GRBs with incomplete dataset, 
we found that the number of XRFs, XRRs, and GRBs are 20, 126, and 60 respectively.  
The numbers of XRFs, XRRs, and GRBs in the HETE-2 sample are 16, 19, and 10 
respectively.  Therefore, the ratio of the numbers of XRRs and GRBs is identical 
for both BAT and HETE-2 sample.  
The small numbers of XRFs in the BAT sample is due to the difficulty in observing
very soft XRFs in the BAT \citep{band2003,band2006}.  
However, as mentioned in \citet{band2006}, it is
very difficult to determine the actual detection threshold of the BAT due to its 
complexity in the triggering algorithm.  Although nothing could be addressed about
the actual number of XRFs, the number of GRBs in XRRs and GRBs seen in
the BAT sample is consistent with the HETE-2 sample.  The detailed study of the 
Swift XRFs and XRRs is presented elsewhere \citep{swift_xrfxrr}.  

\begin{figure}
\centerline{
\includegraphics[scale=1.0,angle=0]{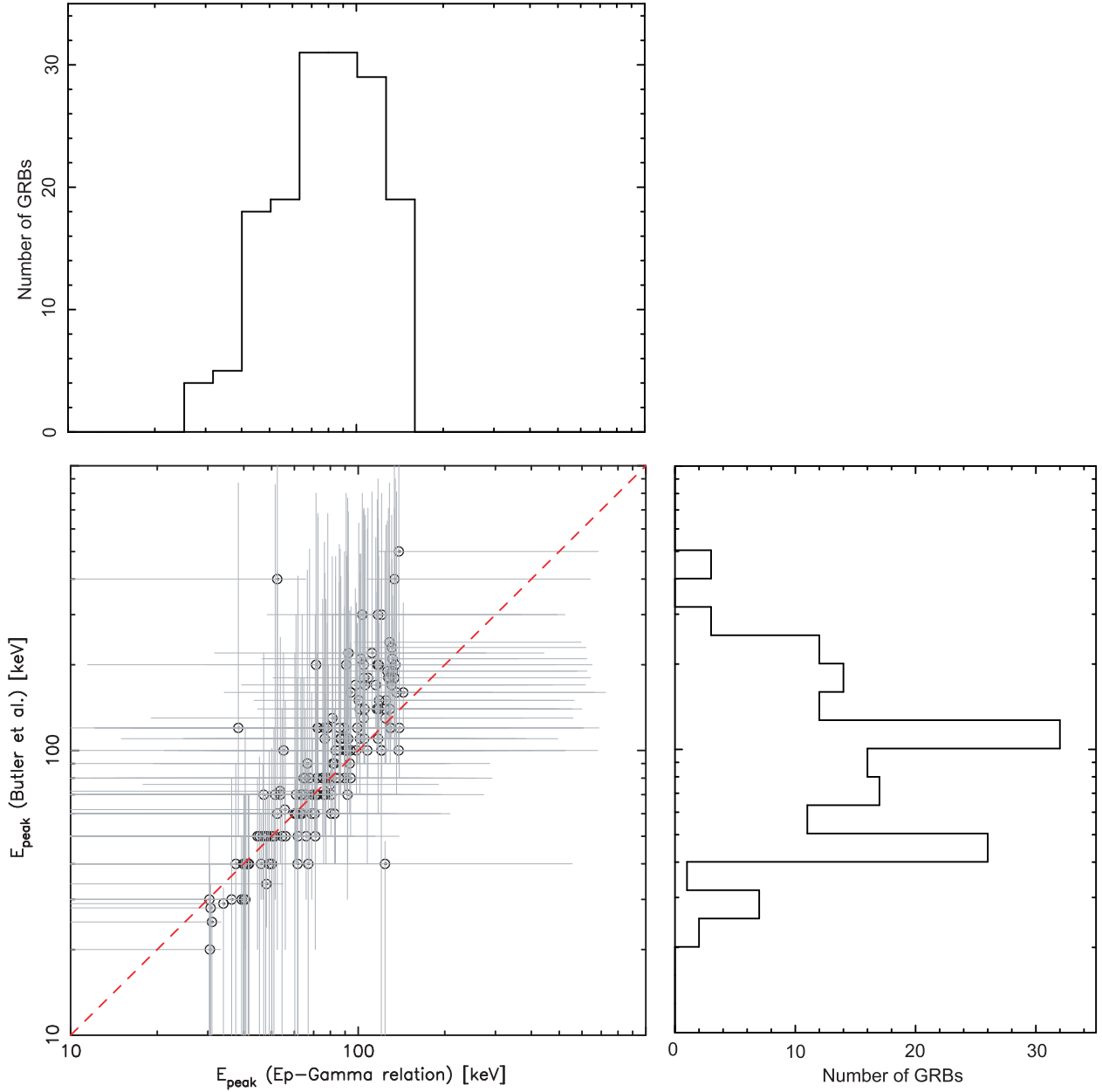}}
\caption{The relationship between $\ep$ reported by \citet{butler2007}
and $\ep$ derived from the weighted $\ep$ - $\Gamma$ relation for the
Band function.  The sample only contains long bursts which have a PL 
photon index $\Gamma$ from 1.3 to 2.3.  Both $\ep$ distributions of 
\citet{butler2007} and the $\ep$ - $\Gamma$ relation for the same sample 
are shown in the histograms.}
\label{fig:comp_taka_nat_ep}
\end{figure}

\citet{butler2007} calculated $\ep$ by their Bayesian approach for 218 {\it Swift} 
GRBs using only the BAT data.  Based on their calculated $\ep$ and the bolometric fluence, 
they claimed that all of the empirical relations, $\eps$ - $\eiso$ \citep{amati2002}, 
$\eps$ - $\lpiso$ \citep{yonetoku2004}, and $\eps$$T_{45}$ - $\lpiso$ \citep{firmani2006}, 
proposed in the pre-{\it Swift} observations are not valid for the {\it Swift} BAT sample.  
We investigated the validity of their $\ep$ by checking $\ep$ obtained by using our 
$\ep$ - $\Gamma$ relation.  
We created the BAT spectra for their GRB samples by the time interval 
reported on Table 1 of \citet{butler2007}.  Then, we fit the spectrum by a PL 
model to extract the best fit $\Gamma$.  By only selecting their $\Gamma$ within 
the allowed $\Gamma$ range for applying our $\ep$ - $\Gamma$ relation ($1.3 < \Gamma 
< 2.3$) and also excluding the short GRBs (156 samples in total), 
we calculated $\ep$ applying our weighted $\ep$ - $\Gamma$ relation for 
the Band function.  
Figure \ref{fig:comp_taka_nat_ep} shows the $\ep$ reported on 
\citet{butler2007} versus $\ep$ derived from our weighted $\ep$ - $\Gamma$ relation for 
the Band function.  
Although the error bars are large in both estimators, 
the figure shows that $\ep$ of the \citet{butler2007} sample 
has a systematically higher $\ep$ compared to that from our $\ep$ - $\Gamma$ relation.  
About 20\% of the \citet{butler2007} sample selected based on the 
range of $\Gamma$ from 1.3 to 2.3 exceeds $\ep$ $\sim$ 150 keV which is the limit 
of the estimated $\ep$ using our $\ep$ - $\Gamma$ relation for $\Gamma$ = 1.3.
Furthermore, we are already excluding 20\% of the \citet{butler2007} sample because 
those bursts fall 
outside  limit range of $\Gamma$ from 1.3 to 2.3 in our relation.  This 
limit is determined because $\ep$ is very likely located outside of the BAT energy 
range, and therefore, the BAT data alone can not constrain about $\ep$ (the BAT 
data only can provide the limit in $\ep$).  In total, about 35\% of the \citet{butler2007} 
samples are obviously inconsistent with the $\ep$ estimated based on our 
$\ep$ - $\Gamma$ relation.
However, $\ep$ is {\it constrained} in the most of 
the \citet{butler2007} sample.  These results provide a caution for the method for 
estimating $\ep$ in \citet{butler2007}.  
%

\begin{figure}
\centerline{
\includegraphics[scale=0.5,angle=0]{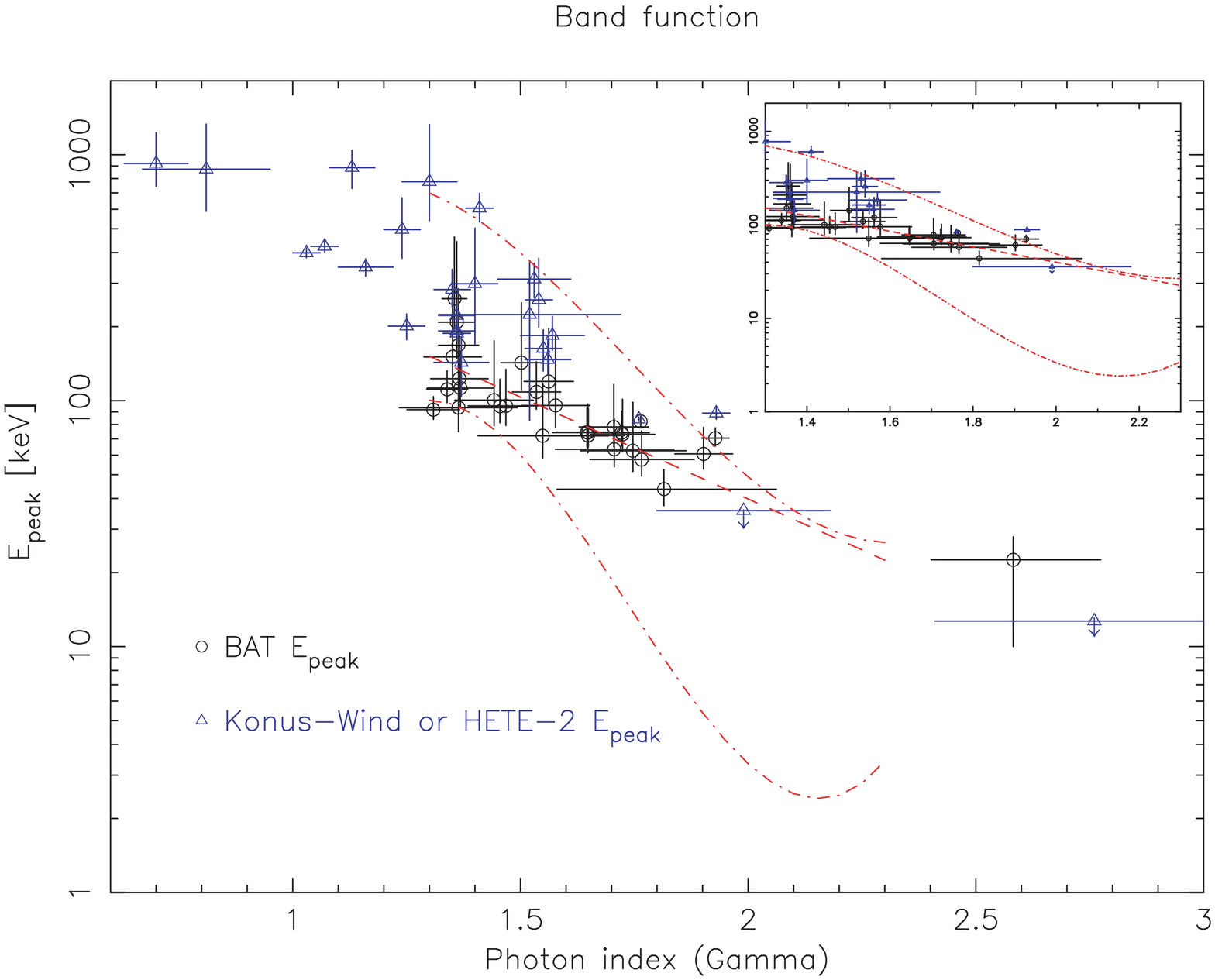}}
\vspace{1cm}
\centerline{
\includegraphics[scale=0.5,angle=0]{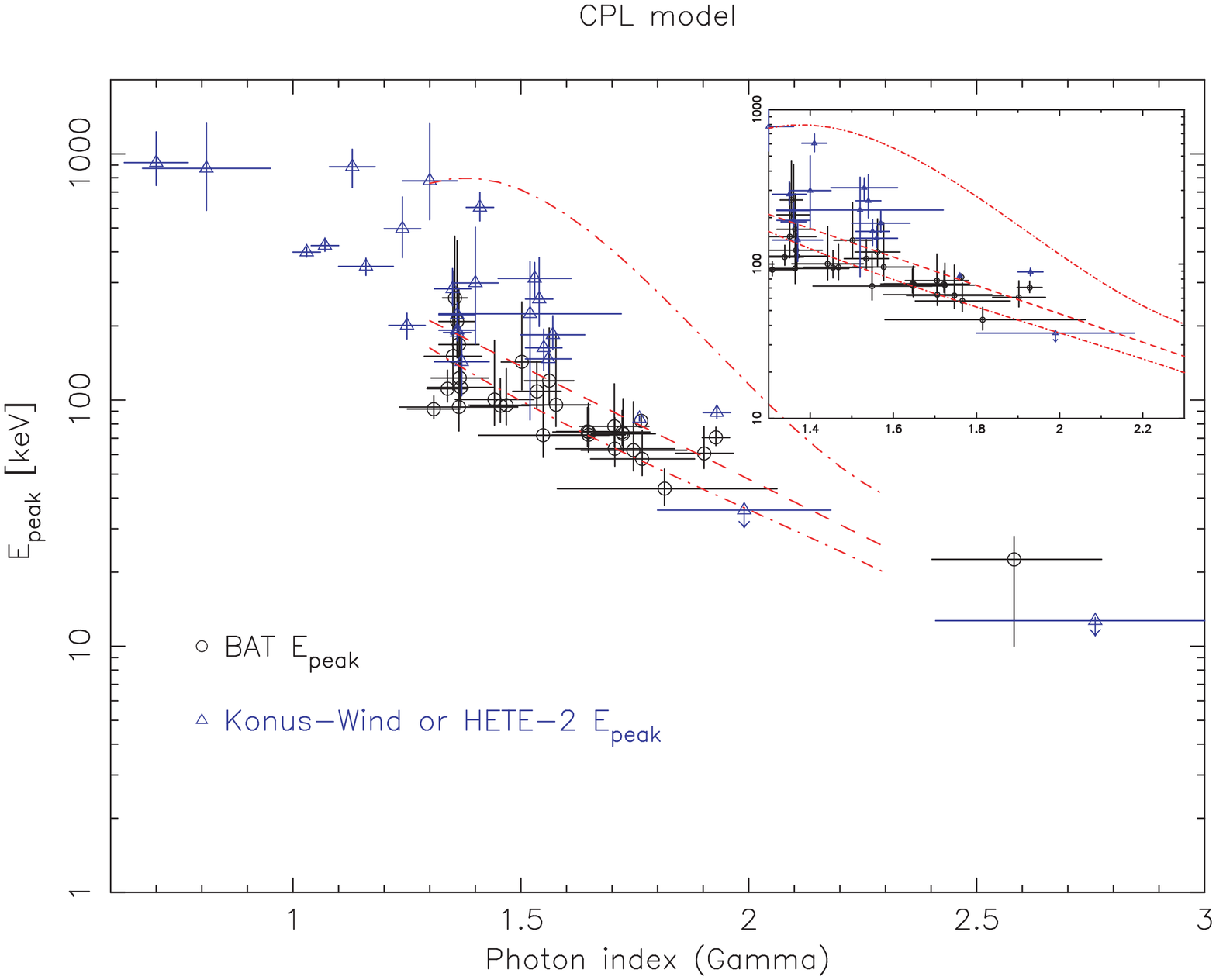}}
\caption{The distribution of $\ep$ and photon index, $\Gamma$, in 
a PL fit in the BAT GRB sample (black circles).  The GRBs which were
 simultaneously observed by {\it HETE-2} and {\it Konus-Wind} are
 overlaid (blue triangles).  The weighted $\ep$ - $\Gamma$ relation 
for the Band function (top) and a CPL model (bottom) with 1-$\sigma$ 
confidence level is overlaid on the data.  {\it Inset}: The extended
 figures of $\Gamma$ from 1.3 to 2.3 where the $\ep$ - $\Gamma$ relation
 is valid.}
\label{fig:comp_epeak_gamma_obs}
\end{figure}

\citet{butler2007} justify their $Swift$-only $\ep$ estimates in part by 
comparing to {\it Konus-Wind} measurements of $\ep$ for the same bursts.  They also 
using the $\ep$ distribution as a prior.  
However, the assumption of $\ep$ measured by {\it Konus-Wind} should be identical 
to that of BAT in \citet{butler2007} might not be valid.  Because BAT has a significantly 
larger effective area and also relatively softer energy band than {\it Konus-Wind}, the 
time interval for creating the time-averaged spectrum based on the BAT data could be 
systematically longer than that of {\it Konus-Wind} (Sakamoto et al. in preparation).  
This longer time interval for the time-averaged spectrum in BAT might lead to a 
systematically lower $\ep$ which might contradict with the $\ep$ based on the {\it Konus-Wind} 
data alone.  
For instance, GRB 060117, which has individual measurements of $\ep$ from the BAT 
and the {\it Konus-Wind} data (see Table 2 and Table 3), shows a smaller 
$\ep$ in the BAT data.  The duration reported based on the {\it Konus-Wind} data is 
$\sim$ 20 seconds \citep{golenetskii2006b}.  On the other hand, the duration used to accumulate 
the BAT spectrum is $\sim$ 30 seconds \citep{sakamoto2007}.  We confirmed based on 
our cross-calibration work that there is no systematic difference in $\ep$ of this burst 
between BAT and {\it Konus-Wind} if we select exactly the same time interval for accumulating 
the spectrum (Sakamoto et al. in preparation).  Therefore, we believe that a prior assumption 
of $\ep$ based on a particular GRB instrument might introduce an another level of a systematic 
error in the analysis.
Most importantly, we believe that testing these 
empirical relations, which require the broad-band spectral properties of the prompt GRB 
emission, by using only the BAT narrow-band data could lead to a wrong conclusion.  Current on-going 
activity for analyzing the spectral data of simultaneously observed BAT GRBs by other 
GRB missions such as {\it Konus-Wind} and {\it Suzaku}/WAM (Sakamoto et al. in 
preparation; Krimm et al. in preparation) is indeed a necessary step to answer for 
the validation of these empirical relations.  
We might want to emphasize that the 1-$\sigma$ confidence level of our $\ep-\Gamma$ 
relation based on the Band function includes most of $\ep$ reported by other 
instruments (see Figure \ref{fig:comp_epeak_gamma_obs}).  Therefore, the confidence level which we 
are quoting in our estimator is large enough to include the systematic problem  
in $\ep$ among the different instruments.  

We report the correlation between $\ep$ and the photon index, $\Gamma$, 
of the BAT prompt emission spectrum based on our simulation study.  
Using this relation, it is possible to estimate $\ep$ from $\Gamma$ in the 
range from 1.3 to 2.3.  
We also performed the spectral simulations for assuming various incident 
angles (0$^{\circ}$, 15$^{\circ}$, 30$^{\circ}$ and 50$^{\circ}$) and 
different spectral models (Band function and CPL).  However, none of these 
systematic effects changes the relation.   
In the application, the $\ep$ - $\Gamma$ relation based on the Band function 
as a prior is the appropriate formula to use.  
The $\ep$ - $\Gamma$ relation could be informative 
for classifying 
the BAT GRBs from the photon index alone as derived from a simple power-law 
model which is the best fit for about 80 \% of the whole population of the 
BAT GRBs.  

\acknowledgements
We would like to thank the anonymous referee for comments and suggestions 
that materially improved the paper.

\clearpage

\begin{deluxetable}{ccccc}
\tabletypesize{\footnotesize}
\rotate
\tablecaption{$\ep$ - $\Gamma$ relation based on the Band function\label{tbl:ep_gamma_band}}
\tablewidth{0pt}
\tablehead{
\colhead{$\theta$} &
\colhead{} &
\colhead{$\ep-\Gamma$ relation} & 
\colhead{1-$\sigma$ lower limit} &
\colhead{1-$\sigma$ upper limit}
}
\startdata
 0 (on-axis)& $\log \ep =$ & $3.312 - 0.817\Gamma$ & $-29.450 + 57.904\Gamma -34.337\Gamma^{2} + 6.445\Gamma^{3}$ &
$-1.073 + 9.840\Gamma -7.065\Gamma^{2} + 1.413\Gamma^{3}$\\
15 & $\log \ep =$ & $3.184 - 0.793\Gamma$ & $-31.986 + 62.511\Gamma -37.070\Gamma^{2} + 6.975\Gamma^{3}$ &
$-1.991 + 11.452\Gamma -7.988\Gamma^{2} + 1.587\Gamma^{3}$\\
30 & $\log \ep =$ & $3.231 - 0.819\Gamma$ & $-20.684 + 43.646\Gamma -26.891\Gamma^{2} + 5.185\Gamma^{3}$ &
$-6.762 + 19.192\Gamma -12.065\Gamma^{2} + 2.291\Gamma^{3}$\\
50 & $\log \ep =$ & $3.210 - 0.796\Gamma$ & $6.782 - 3.948\Gamma -0.286 \Gamma^{2} + 0.348\Gamma^{3}$ &
$-6.860 + 18.110\Gamma -10.740\Gamma^{2} + 1.935\Gamma^{3}$\\
Weighted & $\log \ep =$ & $3.258 -0.829 \Gamma$ & $-20.684 + 43.646\Gamma -26.891\Gamma^{2} + 5.185\Gamma^{3}$ & 
$-5.198 + 16.568\Gamma -10.630\Gamma^{2} + 2.034\Gamma^{3}$\\
\enddata
\end{deluxetable}

\begin{deluxetable}{ccccc}
\tabletypesize{\footnotesize}
\rotate
\tablecaption{$\ep$ - $\Gamma$ relation based on a CPL model\label{tbl:ep_gamma_cpl}}
\tablewidth{0pt}
\tablehead{
\colhead{$\theta$} &
\colhead{} &
\colhead{$\ep-\Gamma$ relation} & 
\colhead{1-$\sigma$ lower limit} &
\colhead{1-$\sigma$ upper limit}
}
\startdata
 0 (on-axis)& $\log \ep =$ & $3.722 - 1.033\Gamma$ & $1.829 +1.874\Gamma - 1.638\Gamma^{2} + 0.315\Gamma^{3}$ &
$-14.504 + 31.357\Gamma -18.043\Gamma^{2} + 3.234\Gamma^{3}$\\
15 & $\log \ep =$ & $3.657 - 0.994\Gamma$ & $1.829 + 1.874\Gamma - 1.638\Gamma^{2} +0.315\Gamma^{3}$ &
$-14.504 + 31.357\Gamma -18.043\Gamma^{2} + 3.234\Gamma^{3}$\\
30 & $\log \ep =$ & $3.490 - 0.904\Gamma$ & $1.980 + 1.434\Gamma -1.342\Gamma^{2} + 0.258\Gamma^{3}$ &
$-8.724 + 20.829\Gamma -11.819\Gamma^{2} + 2.052\Gamma^{3}$\\
50 & $\log \ep =$ & $3.664 - 0.984\Gamma$ & $-0.742 + 5.847\Gamma -3.751 \Gamma^{2} + 0.695\Gamma^{3}$ &
$-1.794 + 8.489\Gamma -4.823\Gamma^{2} + 0.813\Gamma^{3}$\\
Weighted & $\log \ep =$ & $3.518 -0.920 \Gamma$ & $5.018 -3.548\Gamma +1.366\Gamma^{2} - 0.229\Gamma^{3}$ & 
$-9.443 + 22.037\Gamma -12.478\Gamma^{2} + 2.168\Gamma^{3}$\\
\enddata
\end{deluxetable}

\begin{table}
\caption{The BAT time-averaged spectral parameters fitted with a simple power-law 
(PL) model and a power-law times exponential cutoff (CPL) model.  See the BAT1 
catalog paper for the details about the BAT analysis (Sakamoto et
 al. 2008a).  The degree of freedom in a PL fit and a CPL fit is all 57 
and 56 respectively.}
\vspace{0.5cm}
\centerline{
\label{hyo:spec_para}
{\small
\begin{tabular}{cccc|ccc}\\\hline\hline
   & & \multicolumn{2}{c}{PL} & \multicolumn{3}{c}{CPL}\\\cline{3-7}
GRB  & Trigger ID & $\Gamma$ & $\chi^{2}$  & $\alpha$ & $\ep$ [keV] &
 $\chi^{2}$ \\\hline
GRB 041217 & 100116 & $1.46 \pm 0.07$ & 74.9 & $-0.7 \pm 0.3$ & $95_{-14}^{+27}$ & 54.8\\
GRB 041224 & 100703 & $1.72 \pm 0.06$ & 56.1 & $-1.1 \pm 0.3$ & $74_{-9}^{+16}$ & 36.7\\
GRB 050117 & 102861 & $1.50 \pm 0.04$ & 38.8 & $-1.2 \pm 0.2$ & $143_{-33}^{+108}$ & 29.6\\
GRB 050124 & 103647 & $1.47 \pm 0.08$ & 58.7 & $-0.7 \pm 0.4$ & $95_{-16}^{+39}$ & 45.4\\
GRB 050128 & 103906 & $1.37 \pm 0.07$ & 59.3 & $-0.7 \pm 0.3$ & $113_{-19}^{+46}$ & 44.8\\
GRB 050219A & 106415 & $1.31 \pm 0.06$ & 103.2 & $-0.1 \pm 0.3$ & $92_{-8}^{+12}$ & 45.5\\
GRB 050219B & 106442 & $1.53 \pm 0.05$ & 86.6 & $-1.0_{-0.2}^{+0.3}$ & $108_{-16}^{+35}$ & 69.0\\
GRB 050410 & 114299 & $1.65 \pm 0.08$ & 78.5 & $-0.8 \pm 0.4$ & $74_{-9}^{+19}$ & 61.3\\
GRB 050416B & 114797 & $1.4 \pm 0.1$ & 67.4 & $-0.4_{-0.6}^{+0.7}$ & $94_{-19}^{+66}$ & 59.7\\
GRB 050525A$^{a}$ & 130088 & 1.76  & 166.4 & $-1.0 \pm 0.1$ & $82_{-3}^{+4}$ & 17.9\\
GRB 050716 & 146227 & $1.37 \pm 0.06$ & 52.5 & $-0.8 \pm 0.3$ & $123_{-24}^{+61}$ & 39.4\\
GRB 050815 & 150532 & $1.8 \pm 0.2$ & 75.6 & $0.9_{-1.4}^{+1.9}$ & $44_{-6}^{+9}$ & 62.1\\
GRB 050820B & 151334 & $1.34 \pm 0.04$ & 89.6 & $-0.6 \pm 0.2$ & $111_{-13}^{+21}$ & 48.7\\
GRB 050915B & 155284 & $1.90 \pm 0.06$ & 55.5 & $-1.4 \pm 0.3$ & $61_{-8}^{+17}$ & 46.0\\
GRB 051021B & 160672 & $1.6 \pm 0.1$ & 56.9 & $-0.6_{-0.6}^{+0.8}$ & $72_{-13}^{+45}$ & 49.7\\
GRB 060111A & 176818 & $1.65 \pm 0.07$ & 69.0 & $-0.9 \pm 0.3$ & $74_{-10}^{+19}$ & 50.4\\
GRB 060115 & 177408 & $1.8 \pm 0.1$ & 52.6 & $-1.0_{-0.5}^{+0.6}$ & $63_{-11}^{+36}$ & 45.8\\
GRB 060117 & 177666 & $1.93 \pm 0.03$ & 67.0 & $-1.5 \pm 0.1$ & $70_{-5}^{+7}$ & 35.6\\
GRB 060204B & 180241 & $1.44 \pm 0.09$ & 47.0 & $-0.8 \pm 0.4$ & $100_{-21}^{+75}$ & 38.9\\
GRB 060206 & 180455 & $1.71 \pm 0.08$ & 64.6 & $-1.2 \pm 0.3$ & $78_{-13}^{+38}$ & 55.3\\
GRB 060211A & 181126 & $1.8 \pm 0.1$ & 71.5 & $-0.9_{-0.5}^{+0.6}$ & $58_{-8}^{+18}$ & 60.6\\
GRB 060322 & 202442 & $1.58 \pm 0.07$ & 64.6 & $-1.1_{-0.4}^{+0.3}$ & $96_{-18}^{+90}$ & 57.5\\
GRB 060428B & 207399 & $2.6 \pm 0.2$ & 66.7 & $-0.8_{-1.2}^{+1.6}$ & $22_{-13}^{+5}$ & 59.1\\
GRB 060707 & 217704 & $1.7 \pm 0.1$ & 70.5 & $-0.6_{-0.6}^{+0.7}$ & $63_{-10}^{+21}$ & 60.5\\
GRB 060813 & 224364 & $1.36 \pm 0.04$ & 54.1 & $-1.0 \pm 0.2$ & $168_{-39}^{+117}$ & 43.5\\
GRB 060825 & 226382 & $1.72 \pm 0.07$ & 64.7 & $-1.2 \pm 0.3$ & $73_{-11}^{+28}$ & 53.7\\
GRB 060908 & 228581 & $1.35 \pm 0.06$ & 50.7 & $-1.0 \pm 0.3$ & $151_{-41}^{+184}$ & 44.2\\
GRB 060927 & 231362 & $1.65 \pm 0.08$ & 70.4 & $-0.9 \pm 0.4$ & $72_{-11}^{+25}$ & 57.5\\
GRB 070420 & 276321 & $1.56 \pm 0.05$ & 60.7 & $-1.2 \pm 0.2$ & $120_{-24}^{+76}$ & 51.1\\
GRB 070508 & 278854 & $1.35 \pm 0.03$ & 38.4 & $-1.1 \pm 0.1$ & $260_{-68}^{+203}$ & 27.8\\
GRB 070521 & 279935 & $1.36 \pm 0.04$ & 57.5 & $-1.1 \pm 0.2$ & $209_{-60}^{+234}$ & 50.1\\\hline
\end{tabular}}
}
\tablenotetext{a}{The confidence interval is not calculated because of $\chi^{2}_{\nu}$ $>$ 2.}
\tablenotetext{}{Short GRBs, GRB 050820A and GRB 050925 are excluded.}
\end{table}

\begin{table}
\caption{The spectral parameters of simultaneously observed by {\it Konus-Wind} or 
{\it HETE-2}.}
\vspace{0.5cm}
\label{hyo:spec_para_kw_hete2}
\centerline{
{\small
\begin{tabular}{ccccccc}\\\hline\hline
GRB & Model & $\alpha$ & $\beta$ & $\ep$ & $\Gamma_{\rm{BAT}}$ & Reference\\\hline
GRB 050215B & Band & - & - & $<$ 35.7$^{a}$ & $2.0 \pm 0.2$ & Nakagawa et al. (2005)\\
GRB 050326  & Band & $-0.74 \pm 0.09$ & $-2.49 \pm 0.16$ & $201 \pm 24$ & $1.25 \pm 0.04$ & Golenetskii et al. (2005a)\\
GRB 050525A & CPL  & $-1.10 \pm 0.05$ & - & $84.1 \pm 1.7$ & 1.76$^{c}$ & Golenetskii et al. (2005b)\\
GRB 050603  & Band & $-0.79 \pm 0.06$ & $-2.15 \pm 0.09$ & $349 \pm 28$ & $1.16 \pm 0.06$ & Golenetskii et al. (2005c)\\
GRB 050713A & CPL  & $-1.12 \pm 0.08$ & - & $312 \pm 50$ & $1.53 \pm 0.08$ & Golenetskii et al. (2005d)\\
GRB 050824  & Band & - & - & $<$ 12.7$^{b}$ & $2.8 \pm 0.4$ & Crew et al. (2005a)\\
GRB 050922C & CPL  & $-0.83_{-0.23}^{+0.26}$ & - & $143 \pm 39$ & $1.37 \pm 0.06$ & Crew et al. (2005b)\\
GRB 051008  & CPL  & $-0.975_{-0.078}^{+0.086}$ & - & $886 \pm 157$ & $1.13 \pm 0.05$ & Golenetskii et al. (2005e)\\
GRB 051109A & CPL  & $-1.25_{-0.44}^{+0.59}$ & - & $224 \pm 141$ & $1.5 \pm 0.2$ & Golenetskii et al. (2005f)\\
GRB 060105  & CPL  & $-0.83 \pm 0.03$ & - & $424_{-22}^{+25}$ & $1.07 \pm 0.04$ & Golenetskii et al. (2006a)\\
GRB 060117  & Band & $-1.52_{-0.07}^{+0.08}$ & $-2.9_{-0.5}^{+0.3}$ & $89 \pm 5$ & $1.93 \pm 0.03$ & Golenetskii et al. (2006b)\\
GRB 060313  & CPL  & $-0.6 \pm 0.2$ & - & $922_{-177}^{+306}$ & $0.70 \pm 0.07$ & Golenetskii et al. (2006c)\\
GRB 060510  & CPL  & $-1.66 \pm 0.07$ & - & $184_{-24}^{+36}$ & $1.57 \pm 0.07$ & Golenetskii et al. (2006d)\\
GRB 060813  & Band & $-0.53_{-0.14}^{+0.16}$ & $-2.6_{-0.5}^{+0.3}$ & $192_{-18}^{+20}$ & $1.36 \pm 0.04$ & Golenetskii et al. (2006e)\\
GRB 060814  & CPL  & $-1.4 \pm 0.2$ & - & $257_{-58}^{+122}$ & $1.54 \pm 0.03$ & Golenetskii et al. (2006f)\\
GRB 060904A & Band & $-1.0 \pm 0.2$ & $-2.6_{-1.0}^{+0.4}$ & $163 \pm 31$ & $1.55 \pm 0.04$ & Golenetskii et al. (2006g)\\
GRB 061007  & Band & $-0.7 \pm 0.4$ & $-2.6_{-0.5}^{+0.3}$ & $399_{-18}^{+19}$ & $1.03 \pm 0.03$ & Golenetskii et al. (2006h)\\
GRB 061021  & CPL  & $-1.2 \pm 0.1$ & - & $777_{-237}^{+549}$ & $1.30 \pm 0.06$ & Golenetskii et al. (2006i)\\
GRB 061121  & CPL  & $-1.32 \pm 0.05$ & - & $606_{-72}^{+90}$ & $1.41 \pm 0.03$ & Golenetskii et al. (2006j)\\
GRB 061201  & CPL  & $-0.36_{-0.65}^{+0.40}$ & - & $873_{-284}^{+458}$ & $0.8 \pm 0.1$ & Golenetskii et al. (2006k)\\
GRB 061222A & Band & $-0.94_{-0.13}^{+0.14}$ & $-2.4_{-1.2}^{+0.3}$ & $283_{-42}^{+59}$ & $1.35 \pm 0.04$ & Golenetskii et al. (2006l)\\
GRB 070220  & Band & $-1.2_{-0.2}^{+0.3}$ & $-2.0_{-0.4}^{+0.3}$ & $299_{-130}^{+204}$ & $1.40 \pm 0.04$ & Golenetskii et al. (2007a)\\
GRB 070328  & Band & $-1.0 \pm 0.1$ & $-2.0_{-0.4}^{+0.2}$ & $496_{-117}^{+172}$ & $1.24 \pm 0.04$ & Golenetskii et al. (2007b)\\
GRB 070420  & CPL  & $-1.2 \pm 0.2$ & - & $147_{-19}^{+29}$ & $1.56 \pm 0.05$ & Golenetskii et al. (2007c)\\
GRB 070508  & CPL  & $-0.81 \pm 0.07$ & - & $188 \pm 8$ & $1.36 \pm 0.03$ & Golenetskii et al. (2007d)\\
GRB 070521  & CPL  & $-0.9 \pm 0.1$ & - & $222_{-21}^{+27}$ & $1.36 \pm 0.04$ & Golenetskii et al. (2007e)\\\hline
\end{tabular}}
}
\tablenotetext{a}{99\% upper limit}
\tablenotetext{b}{90\% upper limit}
\tablenotetext{c}{The confidence interval is not calculated because of $\chi^{2}_{\nu}$ $>$ 2.}
\end{table}

\newpage

\newpage

\newpage

\newpage

\newpage

\newpage

\newpage

\newpage

\newpage
\begin{figure}
\centerline{
\includegraphics[scale=0.5,angle=-90]{f9a.eps}}
\vspace{1cm}
\centerline{
\includegraphics[scale=0.5,angle=-90]{f9b.eps}}
\caption{The best fit weighted $\ep$ - $\Gamma$ relations by the 
incident angles (solid line) and the 
lower and higher 1-$\sigma$ confidence level of the relations (dashed
 lines) for the Band function (top) and a CPL model (bottom) with the
 data points (circles: $\ep$ at the peak of the histogram of
 $\Gamma$, squares: $\ep$ value of 16\% crossing point of the histogram
 of $\Gamma$, and triangles: $\ep$ value of 84\% crossing point of the histogram
 of $\Gamma$).}
\label{fig:weighted_ep_gamma_fit}
\end{figure}

\newpage

\newpage






\clearpage


\begin{thebibliography}{}
\bibitem[Amati et al.(2002)]{amati2002} Amati, L., et al. 2002, A\&A,
			    390, 81
\bibitem[Amati(2003)]{amati2003} Amati, L., 2003, ChJAA, Vol. 3, Supplement, pp.
 455-460
\bibitem[Barthelmy et al.(2005)]{barthelmy2005} Barthelmy, S.D., et al. 2005, 
        Space Sci. Rev., 120, 143
\bibitem[Band et al.(1993)]{band1993}
        Band, D. L., et al. 1993, \apj, 413, 281
\bibitem[Band (2003)]{band2003}
        Band, D. L., 2003, ApJ, 588, 945
\bibitem[Band (2006)]{band2006} 
	Band, D. L., 2006, ApJ, 664, 378 
\bibitem[Barraud, et al.(2005)]{barraud2005}
	Barraud, C., Daigne, F., Mochkovitch, R., Atteia, J. L. 2005,
			    A\&A, 440, 809
\bibitem[Butler et al.(2007)]{butler2007}
	Butler, N.R., Kocevski, D., Bloom, J.S., Curtis, J.L. 2007, \apj, 671, 656
\bibitem[Crew et al.(2005a)]{crew2005a} Crew, G., et al. 2005a, GCN Circ. 3890, http://gcn.gsfc.nasa.gov/gcn3/3890.gcn3
\bibitem[Crew et al.(2005b)]{crew2005b} Crew, G., et al. 2005b, GCN Circ. 4021, http://gcn.gsfc.nasa.gov/gcn3/4021.gcn3
\bibitem[Dermer et al.(1999)]{dermer1999} 
        Dermer, C. D., Chiang, J., and B$\ddot{\rm o}$ttcher 
        1999, \apj, 513, 656 
\bibitem[Dermer and Mitman(2003)]{dermer2003} 
        Dermer, C. D., and Mitman, K. E. 2003, in ASP Conf. Ser. 312, Third 
	Rome Workshop on Gamma-Ray Bursts in the Afterglow Era, ed. M. Feroci 
	et al. (San Francisco: ASP), 301
\bibitem[Firmani et al.(2006)]{firmani2006} 
	Firmani, C, Ghisellini, G., Avila-Reese, V., and Ghirlanda, G. 
	2006, MNRAS, 370, 185
\bibitem[Gehrels et al (2004)]{gehrels2004} Gehrels, N., et al. 2004,
			    ApJ, 611, 1005

\bibitem[Ghirlanda et al.(2004)]{ghirlanda2004} Ghirlanda, G.,
                            Ghisellini, G., Lazzati, D., 2004, ApJ, 616, 331
\bibitem[Golenetskii et al.(2005a)]{golenetskii2005a} Golenetskii, S., et al. 2005a, GCN Circ. 3152, http://gcn.gsfc.nasa.gov/gcn3/3152.gcn3
\bibitem[Golenetskii et al.(2005b)]{golenetskii2005b} Golenetskii, S., et al. 2005b, GCN Circ. 3474, http://gcn.gsfc.nasa.gov/gcn3/3474.gcn3
\bibitem[Golenetskii et al.(2005c)]{golenetskii2005c} Golenetskii, S., et al. 2005c, GCN Circ. 3518, http://gcn.gsfc.nasa.gov/gcn3/3518.gcn3
\bibitem[Golenetskii et al.(2005d)]{golenetskii2005d} Golenetskii, S., et al. 2005d, GCN Circ. 3619, http://gcn.gsfc.nasa.gov/gcn3/3619.gcn3
\bibitem[Golenetskii et al.(2005e)]{golenetskii2005e} Golenetskii, S., et al. 2005e, GCN Circ. 4078, http://gcn.gsfc.nasa.gov/gcn3/4078.gcn3
\bibitem[Golenetskii et al.(2005f)]{golenetskii2005f} Golenetskii, S., et al. 2005f, GCN Circ. 4238, http://gcn.gsfc.nasa.gov/gcn3/4238.gcn3
\bibitem[Golenetskii et al.(2006a)]{golenetskii2006a} Golenetskii, S., et al. 2006a, GCN Circ. 4439, http://gcn.gsfc.nasa.gov/gcn3/4439.gcn3
\bibitem[Golenetskii et al.(2006b)]{golenetskii2006b} Golenetskii, S., et al. 2006b, GCN Circ. 4542, http://gcn.gsfc.nasa.gov/gcn3/4542.gcn3
\bibitem[Golenetskii et al.(2006c)]{golenetskii2006c} Golenetskii, S., et al. 2006c, GCN Circ. 4881, http://gcn.gsfc.nasa.gov/gcn3/4881.gcn3
\bibitem[Golenetskii et al.(2006d)]{golenetskii2006d} Golenetskii, S., et al. 2006d, GCN Circ. 5113, http://gcn.gsfc.nasa.gov/gcn3/5113.gcn3
\bibitem[Golenetskii et al.(2006e)]{golenetskii2006e} Golenetskii, S., et al. 2006e, GCN Circ. 5446, http://gcn.gsfc.nasa.gov/gcn3/5446.gcn3
\bibitem[Golenetskii et al.(2006f)]{golenetskii2006f} Golenetskii, S., et al. 2006f, GCN Circ. 5460, http://gcn.gsfc.nasa.gov/gcn3/5460.gcn3
\bibitem[Golenetskii et al.(2006g)]{golenetskii2006g} Golenetskii, S., et al. 2006g, GCN Circ. 5518, http://gcn.gsfc.nasa.gov/gcn3/5518.gcn3
\bibitem[Golenetskii et al.(2006h)]{golenetskii2006h} Golenetskii, S., et al. 2006h, GCN Circ. 5722, http://gcn.gsfc.nasa.gov/gcn3/5722.gcn3
\bibitem[Golenetskii et al.(2006i)]{golenetskii2006i} Golenetskii, S., et al. 2006i, GCN Circ. 5748, http://gcn.gsfc.nasa.gov/gcn3/5748.gcn3
\bibitem[Golenetskii et al.(2006j)]{golenetskii2006j} Golenetskii, S., et al. 2006j, GCN Circ. 5837, http://gcn.gsfc.nasa.gov/gcn3/5837.gcn3
\bibitem[Golenetskii et al.(2006k)]{golenetskii2006k} Golenetskii, S., et al. 2006k, GCN Circ. 5890, http://gcn.gsfc.nasa.gov/gcn3/5890.gcn3
\bibitem[Golenetskii et al.(2006l)]{golenetskii2006l} Golenetskii, S., et al. 2006l, GCN Circ. 5984, http://gcn.gsfc.nasa.gov/gcn3/5984.gcn3
\bibitem[Golenetskii et al.(2007a)]{golenetskii2007a} Golenetskii, S., et al. 2007a, GCN Circ. 6124, http://gcn.gsfc.nasa.gov/gcn3/6124.gcn3
\bibitem[Golenetskii et al.(2007b)]{golenetskii2007b} Golenetskii, S., et al. 2007b, GCN Circ. 6230, http://gcn.gsfc.nasa.gov/gcn3/6230.gcn3
\bibitem[Golenetskii et al.(2007c)]{golenetskii2007c} Golenetskii, S., et al. 2007c, GCN Circ. 6344, http://gcn.gsfc.nasa.gov/gcn3/6344.gcn3
\bibitem[Golenetskii et al.(2007d)]{golenetskii2007d} Golenetskii, S., et al. 2007d, GCN Circ. 6403, http://gcn.gsfc.nasa.gov/gcn3/6403.gcn3
\bibitem[Golenetskii et al.(2007e)]{golenetskii2007e} Golenetskii, S., et al. 2007e, GCN Circ. 6459, http://gcn.gsfc.nasa.gov/gcn3/6459.gcn3
\bibitem[Huang et al.(2002)]{huang2002}
        Huang, Y. F., Dai, Z. G., and Lu, T. 2002, 
        MNRAS, 332, 735
\bibitem[Kaneko et al.(2006)]{kaneko2006}
	Kaneko, Y. et al. 2006, ApJS, 166, 298
\bibitem[Kippen et al.(2002)]{kippen_xrf_astroph}
        Kippen, R. M., Woods, P. M.,  Heise, J., in't Zand, J., Briggs,
        M. S., \& Preece, R. D. 2002, in Gamma-Ray Bursts and
        Afterglow Astronomy, eds. G. R. Ricker and R. Vanderspek (New
        York: AIP), 244
\bibitem[Lamb, Donaghy \& Graziani(2005)]{lamb2005}
        Lamb, D. Q., Donaghy, T. Q., and Graziani, C. 2005, ApJ, 520,
        335
\bibitem[Liang \& Zhang(2005)]{liang2005} Liang, E., Zhang, B., ApJ, 
	633, 611
\bibitem[M\'esz\'aros et al.(2002)]{mes2002} M\'esz\'aros, P., Ramirez-Ruiz, E., 
	Rees, M. J., Zhang, B., ApJ, 578, 812
\bibitem[Mochkovitch et al.(2003)]{mochkovitch2003}
        Mochkovitch, R., Daigne, F., Barraud, C., \& Atteia, J. L.
        2003, in APS Conf. Ser. 312, Third Rome Workshop on Gamma-Ray Bursts in 
	the Afterglow Era, ed. M. Feroci et al. (San Francisco: ASP), 381
\bibitem[Nakagawa et al.(2005)]{nakagawa2005} Nakagawa, U., et al. 2005, GCN Circ. 3053, http://gcn.gsfc.nasa.gov/gcn3/3053.gcn3
\bibitem[Rossi et al.(2002)]{rossi2002} 
        Rossi, E., Lazzati, D., and Rees, M. J. 2002, MNRAS, 
        332, 945 
\bibitem[Sakamoto et al.(2004)]{sakamoto2004} Sakamoto, T., et al. 2004, ApJ, 602, 875
\bibitem[Sakamoto et al.(2005)]{sakamoto2005} Sakamoto, T., et al. 2005, 
	ApJ, 629, 311
\bibitem[Sakamoto et al.(2006)]{sakamoto2006} Sakamoto, T., et al. 2006, ApJ, 636, L73
\bibitem[Sakamoto et al.(2008a)]{sakamoto2007} Sakamoto, T., et
			    al. 2008a, ApJS, 175, 179
\bibitem[Sakamoto et al.(2008b)]{swift_xrfxrr} Sakamoto, T., et al. 2008b, ApJ, 679, 570
\bibitem[Sato et al.(2005)]{rsato2005} Sato, R., et al. 2005, PASJ, 57, 1031
\bibitem[Toma et al.(2005)]{toma2005} Toma, K., Yamazaki, R., Nakamura, 
			    T. 2005, ApJ, 635, 481
\bibitem[Yamazaki et al.(2004)]{yamazaki2004} Yamazaki, R., Ioka, K., 
			    Nakamura, T. 2004, ApJ, 607, L103
\bibitem[Yonetoku et al.(2004)]{yonetoku2004} Yonetoku, D., et al. 2004, ApJ, 609, 935
\bibitem[Zhang \& M\'esz\'aros(2002)]{zhang2002}
        Zhang, B. \& M\'esz\'aros, P. 2002, \apj, 571, 876
\bibitem[Zhang et al.(2004)]{zhang2004}
        Zhang, B., Dai, X., Lloyd-Ronning, N. M., \& M\'esz\'aros, P.
        2004, 601, L119

\end{thebibliography}
\end{document}